2023

# Modeling tsunami inundation for hazard assessment of the Coral Coast of Fiji:

COMMUNITIES OF SIGATOKA AND CUVU

*Diego Arcas[1,2] and Christopher Moore[1]*
[1]NOAA Center for Tsunami Research/Pacific Marine Environmental Laboratory (PMEL), Seattle, WA
[2]Cooperative Institute for Climate, Ocean and Ecosystems Studies. University of Washington, Seattle, WA

# Table of Contents

# Abstract

A tsunami hazard assessment was conducted for the communities of Sigatoka and Cuvu, located on the island of Viti Levu, Fiji. The study presents an overview of historical seismic and tsunami impact on the Pacific Island nation of Fiji in order to identify source areas of relevance to the study. In addition, the sensitivity of the Coral Coast of Viti Levu to tsunamis coming from different parts of the New Hebrides and Tonga-Kermadec Subduction zones was investigated and used to select worst case tsunami scenarios. Following the methodology developed at the NOAA Center for Tsunami Research (Tang et al, 2006) more than 40 tsunami simulations were conducted on a low-resolution model. Four of the sources simulated were selected as the most hazardous to the area of interest and were subsequently modeled on a high-resolution (10 meter) grid of Sigatoka and Cuvu. To these sources, an ensemble of 20 additional sources in the Tonga-Kermadec Subduction Zone were selected for evaluation. These sources were identified in an Intergovernmental Oceanographic Commission (IOC) report (IOC Working Report No. 289, 2018) by a group of seismic experts on tsunami sources in the region. Finally, one source from a 2022 study conducted by NCTR for the US Department of State to assess tsunami hazard to the city of Suva, Fiji using a probabilistic approach was selected for simulation on the high-resolution model. The non-linear shallow water wave inundation Method Of Splitting Tsunami (MOST) model (Titov and González, 1997) is one of the tsunami modeling codes in use by the National Tsunami Hazard Mitigation Program (NTHMP, 2011), (NTHMP, 2017) of the US to conduct tsunami inundation studies with the purpose of developing evacuations maps. The MOST code was used in this study to determine tsunami maximum amplitude, arrival and duration times, flow depths, tsunami inundation, current speeds and attenuation.

# 1. Introduction

Fiji's low-lying coastal areas are particularly vulnerable to tsunami inundation. These areas are home to a large portion of the population and are also important for tourism and agriculture. In the event of a tsunami, these areas could be flooded by tsunami waves, causing damage to infrastructure, homes, and businesses. To mitigate the risk of tsunami inundation, the government of Fiji has implemented several measures. These include educating the public about tsunami preparedness, building tsunami evacuation routes and shelters, and conducting regular drills to practice evacuation procedures (Prevention Web). Fiji's Department of Mineral and Natural Resources also plays a critical role in tsunami warning and response. The Department runs a Tsunami Warning Center which will issue warnings and alerts to the public in case of a tsunami threat (Fiji Museum). In addition, they have set up warning sirens and communication systems to quickly disseminate information to the public in the event of a tsunami. In spite of these efforts, the risk of tsunami inundation remains high for Fiji. Rising sea levels in the South Pacific can also exacerbate the impact of a tsunami. In conclusion, tsunami inundation is a significant threat to Fiji and its people. The government, along with other organizations, has taken steps to mitigate the risk, but the potential for damage and loss of life remains high. It is important for the government to continue to take steps to reduce the risk and to educate the public about tsunami preparedness. This report is one more in a series of measures already taken by the government of Fiji to mitigate tsunami risk.

# 2. Background

The oceans provide world-wide food security and are the economic driver of economies around the globe. From fisheries to energy production, tourism, and goods transport, millions of people rely on the oceans for their livelihood. The ocean and its inhabitants, however, are under increasing pressure from a number of natural disasters such as warming planet, acidification, pollution, population growth, and diminished life forms that support a fishing stock food chain, earthquakes and tsunamis.

Recognizing the need for action, the United Nations declared 2021-2030 the UN Decade of Ocean Science for Sustainable Development to bring attention and resources to efforts that "turn scientific knowledge and understanding into effective actions supporting improved ocean management, stewardship, and sustainable development" (U.N. Decade). Underpinning an ambitious implementation plan is a hierarchy of challenges, programs, projects, actions, and outcomes developed through peer review that together provide a roadmap to the 'ocean we want' by the year 2030. At the top tier, stakeholders are challenged to identify and take actions that contribute to desired societal outcomes. Actions may take the form of topical programs, standalone projects, or specific activities.

The work presented here directly supports an activity of the 'The Ocean Decade Tsunami Program' to implement the domestically successful Tsunami Ready Program in Pacific Ocean Island countries vulnerable to tsunami waves. Societal outcomes are "A safe ocean" with the goal of protecting communities from ocean hazards and "A transparent and accessible Ocean" aimed at building decision making capacity through the sharing of data, information, and technology. In the broadest sense, this topical program is meeting the Challenge to 'Increase community resilience to ocean hazards' by providing communities with the capacity to develop and implement tsunami mitigation strategies.

## 2.1 Tsunami History and Tectonic Setting of the Fiji Islands

A query of the USGS historical earthquake database (USGS) for seismic events since 1700 with magnitude larger than M7.0 in the region surrounding the Fiji Islands, returns more than 65 events mainly along the Tonga-Kermadec and New Hebrides trench. Eleven of those events exceed M8.0 with one of the largest magnitude events M8.2 having occurred as recently as August 19, 2018, 267 km east of Levuka. These figures reveal an active seismic environment surrounding the Fiji Islands with the potential to generate destructive tsunamis along the coast. A similar search of the National Center for Environmental Information historical tsunami database reveals more than 30 significant regional and distant tsunami events with measurable runup along the Fijian coastline.
The location of the island Nation of Fiji between two major subduction zones and the intense seismic and tsunami activity around the islands in recent history elicit the development of a tsunami hazard assessment for the country's coastline.

## 2.1.1. Tectonic Setting

Due to its geographical location in close proximity to two major subduction zones, tsunami inundation is a significant threat to the Pacific Island nation of Fiji. The country is located in the northeast corner of the Indo-Australian Plate near where it subducts under the Pacific Plate. It sits between two of the fastest converging subduction zones, the Tonga-Kermadec (53-85 mm/yr) (Billen and Stock, 2000) to the east and the New Hebrides trench (118 mm/yr) (Richards et al. 2011) to the west. Both these trenches are responsible for the region's high seismic activity and frequent tsunamigenic earthquakes. The Hunter Fracture Zone (See Figure 1) running immediately south of the Fiji Islands and connecting with the southern end of the New Hebrides trench, while located in very close proximity to Fiji, it is mostly inactive (Begg and Gray, 2002) with only a few events in recent history characterized by predominantly lateral tectonic motion resulting in mildly tsunamigenic seismic events dominated by strike-slip faulting.

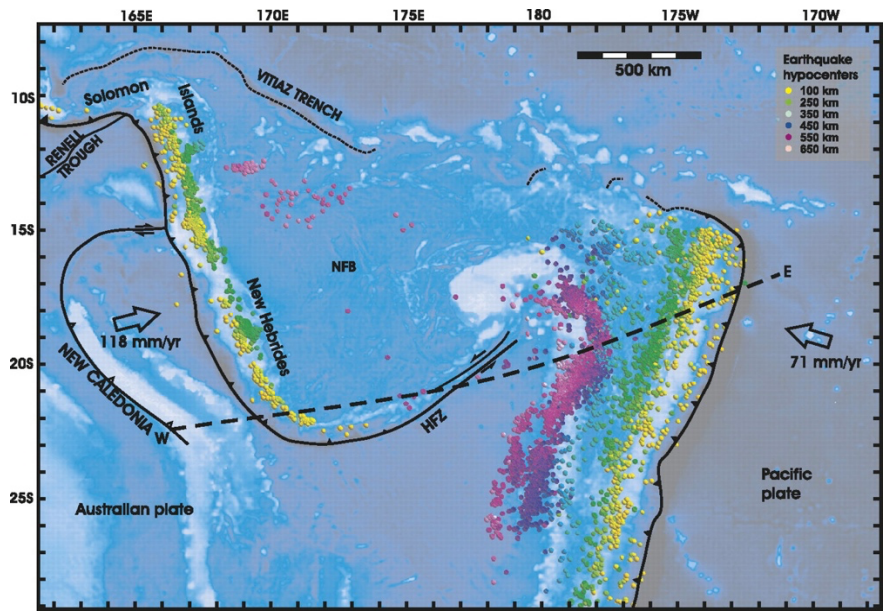

*Figure 1: Topography, bathymetry, and major tectonic element map of the study area. The Tonga and Vanuatu subduction systems are shown together with the locations of earthquake epicenters. Earthquakes between 0 and 70 km depth have been removed for clarity. Remaining earthquakes are color-coded according to depth. Earthquakes located at 500–650 km depth beneath the North Fiji Basin are also shown. Plate motions for Vanuatu are from the U.S. Geological Survey, and for Tonga from Beavan et al. (2002) (see text for details). Dashed line indicates location of cross section shown in Figure 3. NFB—North Fiji Basin; HFZ—Hunter Fracture Zone. (From Richards et al., 2011))*

### 2.1.2. Tsunami History

According to the National Center for Environmental Information (NCEI) tsunami database and a recent NOAA Center for Tsunami Research (NCTR) tsunami hazard assessment for the capital city of Suva (unpublished). Historically, Fiji has been affected by several major tsunamis, including distant tsunamis such as the May 22, 1960 Chile M9.5 earthquake which generated 0.5 m wave runup, the January 14, 1976 M8.0 earthquake also in the Tonga-Kermadec trench, resulting in wave runup of 0.9 m, the June 22, 1977 M7.2 earthquake in the Tonga-Kermadec trench with 0.5 meters of runup in the capital city and the November 26, 1999 M7.5 event in the New Hebrides trench responsible for measured runups of approximately 1 meter in Fiji. The recent March11, 2011 Japan M9.1 earthquake generated waves of approximately 0.23 meters in elevation at King's Wharf in the capital city of Suva.

Despite the frequent impact of tsunamis from regional and distant tsunami sources, the most catastrophic tsunami event experienced in Fiji in recent history occurred on September 14, 1953 from a relatively small seismic event (M6.4) that struck only 25 km south of Suva (Prasad et al., 2000). This large tsunami event is believed to have been caused by an earthquake induced landslide in the submarine canyons off the southeast coast of Viti Levu (Prasad et al., 2000). This type of landslide-generated tsunamis, while typically exhibiting a more localized effect than seismically generated events, have nonetheless, the potential for producing very high amplitude waves. This was the case in 1953 with waves in excess of 5 m (Rynn and Prasaad, 1997) to the north and south of Suva Harbour. According to an NCTR report cited earlier, a total of five people perished from drowning in this event, three in Suva and two in Nakasaleka.

## 2.2 Study Site

Approximately 330 islands conform the Republic of Fiji with a total land area of 18,272 square km (SOPAC Pacific Islands Applied Geoscience Commission) and a population of 915,756 (2023 estimate based on UN Worldometers). The two largest islands are Viti Levu and Vanua Levu. The capital and largest city in Fiji, is Suva, located on the island of Viti Levu with an estimated population of 77,366. Fiji has a rich and diverse cultural heritage. The indigenous Fijian culture is deeply rooted in traditions including music, dance, storytelling, and unique arts and crafts. Indian culture brought by descendants of laborers brought to the island is also prominent, contributing to the cultural diversity of the country. Tourism along with agriculture, including sugar cane and copra production are significant contributors to Fiji's economy. Additionally, Fiji has a developing mining sector, particular in gold and silver. The Coral Coast located along the southern coast of Viti Levu is one of the most popular tourist destinations in the country. It stretches for approximately 80 km (50 miles) along the southern shores of Viti Levu, starting from the Sigatoka River in the west and extending to Pacific Harbor in the east. The Coral Coast offers a range of beautiful beaches that attract large numbers of tourists with Natadola Beach being regarded as one of Fiji's most stunning beaches and Sigatoka Sand Dunes, known for their impressive dune formations. These low-lying coastal areas are essential to the tourist industry of the Coral Coast but are also vulnerable to tsunami impact, as are many of the resorts and tourist accommodations located in their vicinity. A coral reef extends along the coastline which offers some protection from tsunami waves, however openings in the reef line tend to focus the energy of tsunami waves on specific coastal locations that can be particularly hard-hit by the waves. The town of Sigatoka located by the waters of the Sigatoka River serves as the gateway to the Coral Coast, although located more than 4 km upriver, its outskirts extend all the way to the coast and Sigatoka itself could potentially be hit by tsunami waves travelling upriver.

The focus of this study are the communities of Sigatoka and Cuvu immediately to the west of Sigatoka. Figure 2 shows a Google Earth image of the Coral Coast of Fiji and the communities of Sigatoka and Cuvu.

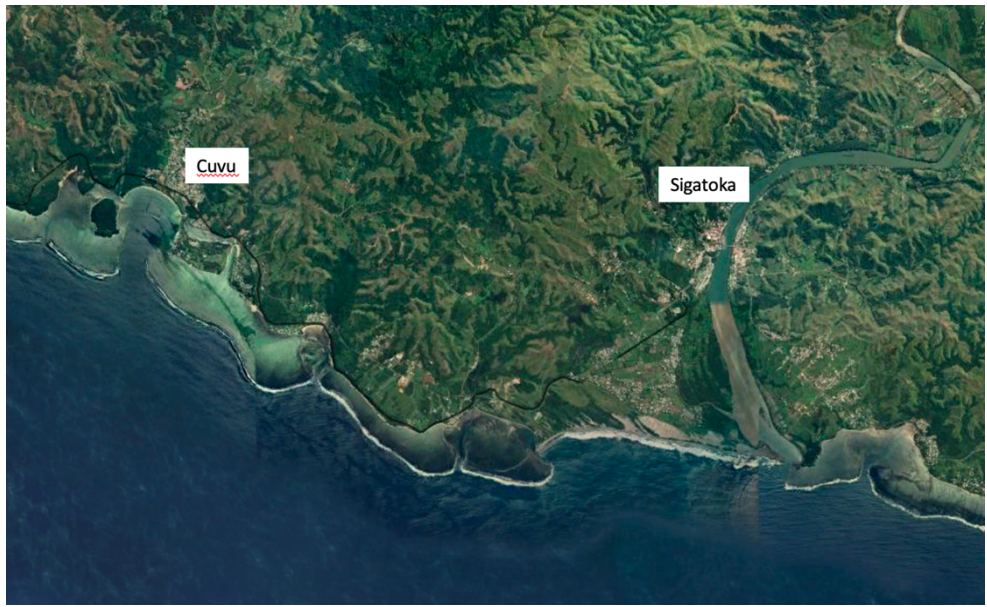

*Figure 2: Google Earth Image of the study area showing the location of the communities of Cuvu to the west and Sigatoka to the east in Fiji's Coral Coast.*

# 3. Inundation Modeling

The general methodology followed in this study for modeling tsunami inundation along coastal areas is to reconstruct tsunami events using a linear combination of seismic unit source propagation solutions in deep water. This reconstruction, in turn, provides boundary and initial conditions to computationally propagate tsunami waves through a set of grids centered at the location of interest and which become finer and finer in resolution the closer tsunami waves get to the area and coastline of concern.

Fine resolution near the area of interest is necessary for two main reasons: High-resolution is necessary to resolve fine coastal features such as bridges, breakwaters, coastal dunes and other local bathymetric and topographic features that have limited geographical extent but may, to a large degree, control inundation. High resolution is also necessary to resolve high-frequency tsunami waves that developed in shallow waters. For accurate calculations, numerical codes require a minimum number of grid nodes per wavelength. Waves with wavelengths too short to satisfy this minimum requirement will experience fast numerical dissipation and will not be modeled accurately. Since tsunami waves experience shortening of its wavelength as they travel from deep onto shallow waters, increasing grid resolution is required in order to preserve the number of grid nodes per wavelength in shallow water. To this effect a set of three nested numerical grids with increasing resolution as the wave approaches the coastline is constructed to be used by the model.

### 3.1 DEM Processing

High resolution topographic and bathymetric data of the area under investigation is necessary to run the numerical model. These data were provided by SPC (Pacific Community) including a high-resolution Light Detection and Radar (LiDAR) dataset of the Coral Coast with precise topographic and bathymetric data in shallow waters. These data sets are essential in the configuration of the computational grids used by the numerical model, but they need to be further processed to reference them to the standard coordinate system used for tsunami modeling (WGS84) and merged with other existing datasets that provide data for areas of the computational grids not covered by the Coral Coast datasets.

Two datasets were provided by SPC (Pacific Community) for this study. A high resolution, 10-meter dataset from a LiDAR survey of the Coral Coast containing topographic and bathymetric data in shallow waters, and a 300-meter resolution dataset of the southwestern region of Viti Levu where the Coral Coast is located (See Figure 3). These datasets were provided in Universal Time Mercator (UTM) projection and were reprojected to WGS84 Coordinate Reference System and elevation values referenced to Mean High Water (MHW). WGS84 and MHW are the standard coordinate reference system and Vertical Datum (VD) used in tsunami hazard studies as it provides a more conservative estimate of tsunami inundation than other lower tidal datums.

While, typically, the transformation of VD involves the development of a 'gradient surface' providing spatial datum variability, for relatively small areas a single offset value can be applied to the entire domain as a good approximation. For this study, a constant offset value of +0.643m between the WGS84 ellipsoid and Mean Sea Level (MSL) was indicated by SPC. Further tidal analysis conducted by SPC based on water level records from Dec. 2013 to Sep. 2014 at Maui Bay on the Coral Coast established offsets between MSL and other tidal datums. Panel b Figure 3 (Courtesy of SPC) shows offsets between the different tidal datums on interest. Based on this information a total elevation correction of 1.22 meters (0.643 m + 0.5835 m) was applied to the original DEM.

Other publicly available datasets were also used to supplement the SPC DEMs in areas not covered by these. For this purpose, two datasets available from NCEI were used, the recently released ETOPO 2022 Global Relief Model (Bedrock Version) which includes data from previous datasets and has been specifically designed to support tsunami forecasting, modeling and warning (NCEI) and the 30-meter resolution multibeam datasets from different bathymetric cruises also available from NCEI (NCEI Multibeam) covering surrounding areas of the Fiji Islands.

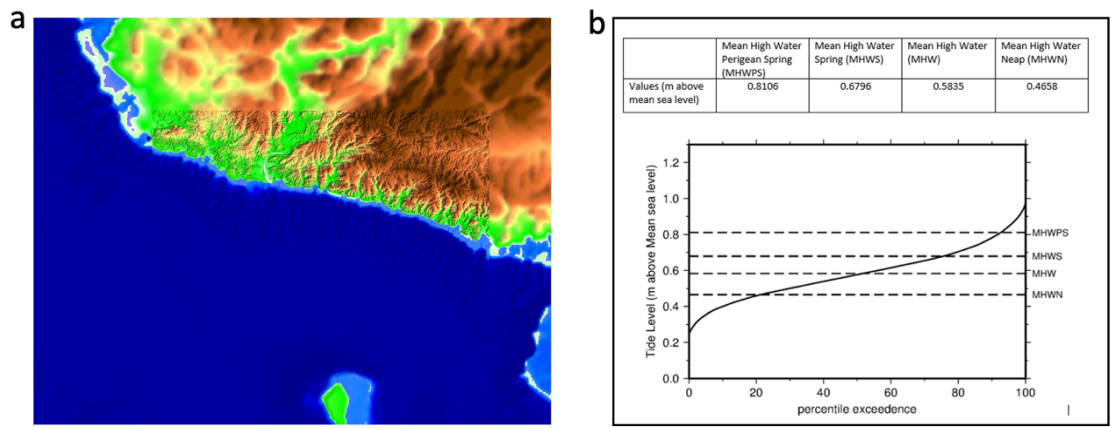

*Figure 3: Panel a) Image representing extent of the topographic and bathymetric dataset provided by SPC. Coverage of the 10-meter bathymetric dataset of the Coral Coast is seen at higher resolution. Panel b) Distance between different tidal datums in the Coral Coast from tice gage analysis provided by SPC.*

## 3.2 Construction of a Base Grid

Once the various datasets have been converted to a common projection and Vertical Datum, a base grid that will be later used to derive the numerical grids was created. This required merging of the different datasets ensuring that low-resolution, less accurate datasets do not contaminate the grid in regions of overlap with high-resolution, more accurate sets. The objective is to provide a base grid with full coverage of all the modeling areas, preserving the highest resolution and most accurate dataset where they exist. The strategy followed in the creation of the base grid is as follows.:

-First a high-resolution mesh of nodes with resolution equal to that of the highest resolution DEM (10 meters) called the base mesh was created. This nodal mesh has large regional coverage which includes the islands of Viti Levu and Vanua Levu. Therefore, some of its grid nodes overlap with those of the 10 meter resolution Coral Coast DEM in this region. The base mesh provides interpolation nodes for the construction of the other numerical grids.

-Next, a new grid (Grid 1) is created by interpolation of the ETOPO2022 dataset onto the base mesh.

-The multibeam data are also interpolated onto the new mesh, creating yet another grid of values (Grid 2). Note that multibeam data are unstructured data with no gridded coverage (Figure 4), therefore data are only interpolated over grid nodes covered by the multibeam.

- The 300-meter resolution SPC dataset of the region around the Coral Coast is interpolated on the interpolation mesh, creating a new grid (Grid 3) with values over the corresponding area of the SPC DEM but with empty (NaN) nodes beyond this region.

- The 10-meter resolution DEM from SPC is interpolated on the base mesh (Grid 4). This interpolation reduces to filling in nodal values in the area covered by the 10-meter SPC with those in the SPC grid since the base mesh was constructed using the SPC-grid nodes as the origin and, as in Grids 2 and 3, empty node values everywhere else. Therefore interpolated data fall on sampled data, preserving the original values.

-Finally, a base grid is created by replacing grid nodes containing data and disregarding nodes containing NaNs, from Grid 2, Grid 3 and Grid 4 onto Grid 1 in that order. This ensures that areas covered by high-resolution more accurate data are not contaminated by other less precise datasets covering the same area, because less accurate data points are always replaced by more accurate ones. Figure 4 shows coverage of the different datasets used in the generation of the base DEM.

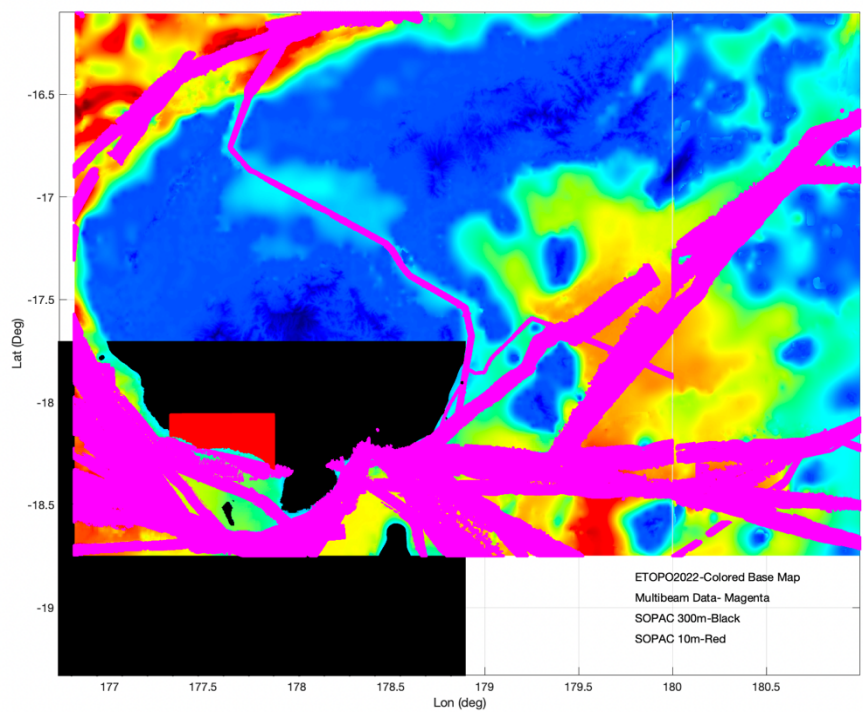

*Figure 4: Map showing areas where the different datasets were used in the construction of the baser grid. Underlying maps has been filled with 15-sec ETOPO2022 data. Multibeam data are show as magenta bands- ship tracks. The black rectangle shows coverage area for the 300-m resolution regional SPC dataset and the red rectangle indicates the coverage area of the 10-m LiDAR survey.*

## 3.3 Numerical Grids

DEM data must be further processed into a set of computational grids that can be used by the numerical model. These should include a set of three nested grids centered around the area of interest, with increasing resolution as they zoom from a large regional domain into the smaller domain of the community under study. The outermost of these three grids is typically labeled as A-grid, the intermediate grid as B-grid and the innermost and highest resolution grid is usually referred to as C-grid. Figure 5 shows geographic coverage of each of the nested grids used in the study. A-grid coverage is represented in blue, B-grid coverage in green and red is used to represent C-grid contiguous coverage for the two communities under study, Cuvu and Sigatoka. In addition to these computational grids precomputed Pacific-wide data retrieved from NCTR's database (Gica et al., 2008) of unit sources was computed using a 4-arcmin grid of the Pacific Ocean.

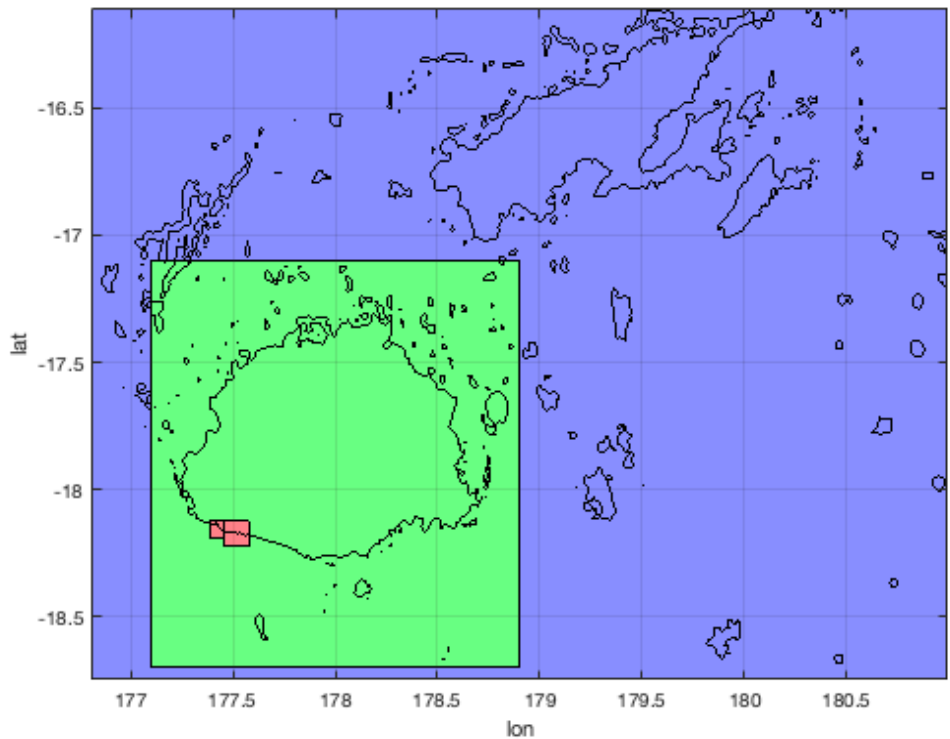

*Figure 5: Areas covered by the set of three nested computational grids used in the study. The blue area shows coverage of the A-grid, the green rectangle indicates coverage of the B-grid and the two red rectangles, the high-resolution, 10-meter grids covering the communities of Sigatoka and Cuvu.*

Construction of the numerical grids is relatively straight-forward once the base grid described in Section 3.2 has been generated. Derivation of the A-, B- and C-grids from the base grid requires only selection of grid extent and resolution, then each of the numerical grids can be generated by simply cropping and subsampling the base grid to the selected values. Table 1 shows the selected values of coverage and resolution for the Sigatoka and Cuvu grids.

*Table 1: Grid statistics for the high-resolution set of grids used in the calculation of the inundation map. Both locations, Sigatoka and Cuvu share grids A and B.*

| Grid | Region | Coverage (Deg) | Cell Res./Size lat x lon (nodes) | CFL |
|---|---|---|---|---|
| | | **High-Resolution Model** | | |
| A | Viti Levu and Vanua Levu | 176.8146°, 180.9951° <br><br> -18.7479°, -16.1132° | 35 arcsec <br><br> 272 x 431 | 5.1163 |
| B | Viti Levu | 177.0993°, 178.9007° <br><br> -18.6993°, -17.0993° | 5 arcsec <br><br> 1153 x 1298 | 0.8520 |
| C-Sigatoka | Sigatoka | 177.4558°, 177.5788° <br><br> -18.2199°, -18.1252° | 1/3 arcsec <br><br> 1046 x 1305 | 0.0979 |
| C-Cuvu | Cuvu | 177.3901°, 177.4556° <br><br> -18.1899°, -18.1181° | 1/3 arcsec <br><br> 793 x 696 | 0.1009 |

Coverage of the A-grid for the study was selected to include the entirety of the islands of Viti Levu and Vanua Levu to ensure that potential wave trapping and inter-island reflections of tsunami waves would be adequately captured by the model using a resolution of 35 arcseconds which should be sufficient to resolve any significant high-frequency waves generated as reflections of neighboring islands.

B-grid coverage was selected to include the entire island of Viti Levu and resolution was selected at 5 arcseconds. This combination of values ensures that a high-resolution grid (5 arcsecond) grid will capture tsunami waves of shortening wavelength as they approach shallower waters of Viti Lebvu from any direction.

Finally, two C-grids were created centered on the communities of Sigatoka and Cuvu with a resolution of 10 meters. Details of the C-grids can be seen in Figure 6. The provided DEM did not include depth values in the Sigatoka River, therefore river depth is set to the offset value of 1.22 meters derived from the correction of the DEM from the WGS84 ellipsoid vertical datum to MHW. Also note, that Lidar data were least accurate immediately to the east of the Sigatoka River mouth compared with other regions of the DEM.

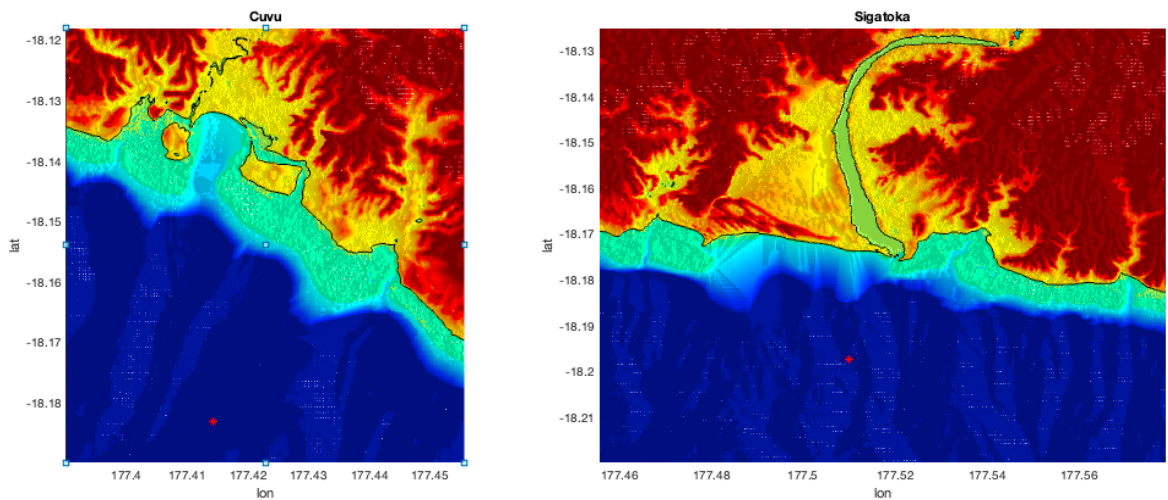

*Figure 6: DEMs of the 10-m resolution grid of the communities of Cuvu (left) and Sigatoka (right). Red dots indicate the location of monitoring points used in source sensitivity studies.*

### 3.4 Source Selection

One of the most important aspects of any tsunami hazard assessment is the selection of tsunami sources that would represent credible worst-case scenarios for the site under study. Two options are exercised in this study: Source compilation from the available literature and systematic source identification from a sensitivity study conducted using the NCTR database.

### 3.4.1 Methodology

In the first option, the existing seismic, tectonic and tsunami literature as well as available databases are consulted. The regional tectonic setting usually indicates the type of rupture mechanism of seismic events. Based on the type of seismic mechanism, seismic sources can be classified according to their tsunamigenic potential. For instance, sources located in areas of active tectonic subduction are likely to exhibit seismic mechanisms with significant vertical dislocation of the ocean bottom, such as reverse and normal thrust faulting. Potential sources located in regions of lateral displacement between tectonic plates will typically present little vertical dislocation of the bottom of the ocean, resulting in moderate tsunamis. In addition, other parameters such as the steepness of the plate interface, the seismic depth of common events, convergence rate between plates and most importantly, the orientation of the trench with respect to the site of interest, reflected in the strike angle, are usually good indicators of the tsunami risk that a tectonic region may have for a particular site. Historical catalogs and databases are also searched in the hope that these will confirm the assumptions made based on tectonic information. However, basing tsunami hazard assessment studies solely on the worst historical events recorded in catalogues may very likely result in an underestimation of the worst credible scenario for the area of concern.

In the present study the high plate convergence rates and seismic mechanisms dominant along the Tonga-Kermadec and New Hebrides subduction zones and the proximity of these trenches to the study site make them the most likely regions for a tsunami worst-case scenario affecting the Fiji Islands. The Hunter Fracture zone to the south of the islands was discarded altogether due to its low seismicity and a predominantly non-tsunamigenic, strike-slip faulting mechanism.

In some cases, the literature can hint more or less explicitly at some of the sources that should be used for worst-case scenario studies, however, these publications are few and far between. To remediate this situation, UNESCOS's Intergovernmental Oceanographic Commission (IOC) has embarked in recent years, in the organization of regional workshops of seismic experts to identify tsunami sources for different purposes including tsunami hazard assessment studies. Fortunately, one of these workshops for the Tonga-Kermadec subduction zone took place in New Zealand in 2018 and the published report clearly identifies worst-case scenario sources for this subduction zone in a manner easily assimilated by tsunami numerical models. Tsunami sources identified in this report was one of the group of sources selected for this study.

No such clarity was found in the literature for the New Hebrides subduction zone, consequently, a different selection approach was employed here. Along this trench, we followed the systematic methodology developed by Tang et al. (2006) that uses NCTR's propagation database (Figure 7) of tsunamis generated by seismic sources. These unit sources are 100 km long by 50 km wide and can be linearly combined to represent tsunamis generated by larger, more complex seismic scenarios. The methodology proceeds as follows: For subduction zones for which seismic ruptures can be long enough for seismic events of magnitude M9 or higher, a magnitude M9.3 event is assumed to be the worst case scenario. This scenario is constructed with uniform slip amount of 25 meters, a length of 1000 km and width of 100 km (i.e. 10 pairs of unit sources), other seismic and tectonic parameters are taken from the NCTR database (Gica et al., 2008) and from the Slab 2.0 (Hayes, G., 2018).

Next, the location and orientation of the event is translated along the length of the subduction zone in steps of 100 km (or the length of one unit source). The full length of the subduction zone is covered in this fashion and all possible orientations of the seismic events are captured.

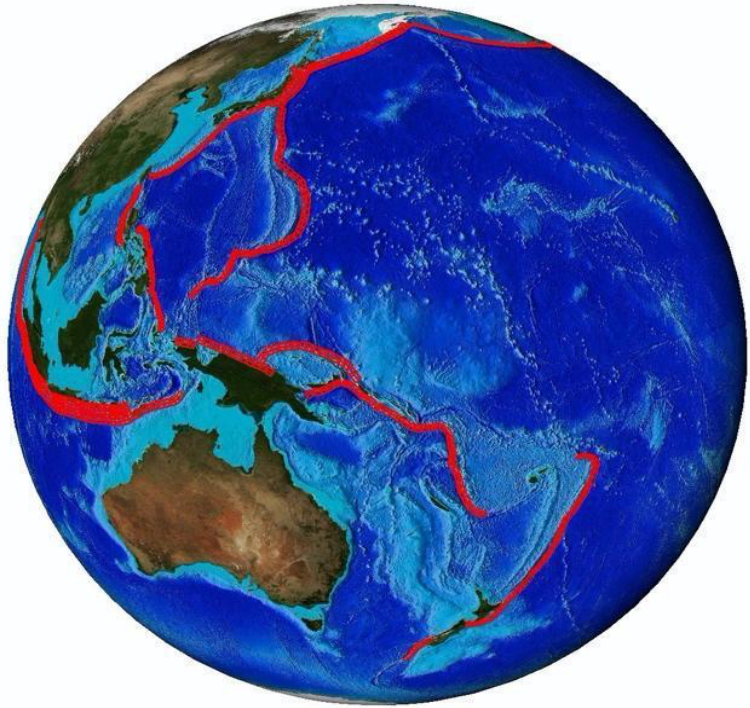

*Figure 7: Map of the Southwestern Pacific Ocean showing the location of the propagation database of unit sources developed at the NOAA Center for Tsunami Research in red.*

Tsunami simulations from each of these scenarios are computed and maximum wave heights at a representative gage point in proximity to the site under study are recorded (See Figure 6). Results of this investigation reveal the sensitivity of the site to tsunamis generated at different locations which facilitates the selection of worst-case scenario events for the study site.

Since this methodology can result in the computation of a large number of scenarios using the high-resolution grids, a less computationally demanding, lower resolution model is developed. This new model has enough coverage and resolution to accurately capture tsunami dynamics in the vicinity of the location of interest, but it does not resolve inundation extents with the accuracy of the higher resolution model described in Section 3.3. For this study two new sets of lower resolution grids (A-, B- and C-grids) were generated for the two sites of interest Sigatoka and Cuvu. Grid extents and resolutions for this new set of grids are shown in Table 2.

*Table 2: Grid statistics for the low-resolution set of grids used in the source sensitivity study. Both locations, Sigatoka and Cuvu share grid A and B.*

| Grid | Region | Coverage (Deg) | Cell Res./Size lat x lon (nodes) | CFL |
|---|---|---|---|---|
| A | Viti Levu and Vanua Levu | 176.8146°, 180.9854° <br><br> -18.7479°, -16.1229° | 105 arcsec <br><br> 91 x 144 | 15.349 |
| B | Viti Levu | 177.0993°, 178.8993° <br><br> -18.6993°, -17.0993° | 15 arcsec <br><br> 385 x 433 | 2.5573 |
| C-Sigatoka | Sigatoka | 177.4558°, 177.5786° <br><br> -18.2198°, -18.1450° | 1 arcsec <br><br> 276 x 435 | 0.2938 |
| C-Cuvu | Cuvu | 177.3901°, 177.4556° <br><br> -18.1899°, -18.1181° | 1 arcsec <br><br> 265 x 232 | 0.3042 |

.

Once worst-case scenarios have been identified following this approach, it is good practice to verify the validity of the low-resolution computations by comparing the results with those obtained after the same scenarios have been run on the high-resolution set of grids. Results of this validation are presented in Section 3.4.2.

Although, the methodology described above was motivated by the lack of clearly defined sources in the literature for the New Hebrides subduction zone, for completeness, the same methodology was applied to the Tonga-Kermadec trench. This resulted in an additional candidate worst-case scenario event being selected in this subduction zone. The approach to investigate the sensitivity of the study sites to different sources using the low-resolution optimized model first, was also applied to the ensemble of sources from the seismic experts report for the Tonga-Kermadec region to identify the most hazardous source in the ensemble.

### 3.4.2 Results of the Source Sensitivity Study

A total of 112 tsunami inundation simulations were run on the optimized models, 56 runs for Sigatoka and 56 for Cuvu. Time series were extracted for each site at representative probe points (Sigatoka 177.5101º, -18.1975º, Cuvu: 177.4144º, -18.1831º) centered in close proximity of the communities of interest and indicated with a red star in Figure 6. Results of these optimized runs are shown in Figures 8 and 9, where the ten largest events from Tonga-Kermadec for the systematic sources, the six largest from New Hebrides and the largest ten from the seismic experts report ensemble are shown for Sigatoka. A similar analysis for Cuvu resulted in the selection of the same worst-case scenarios for this community. This is not surprising given the proximity of the two sites and reinforces the idea that the probe point can be somewhat randomly selected in the domain without any significant impact in the final results. This was further confirmed by the visualization of multiple animations of the simulations which revealed low spatial variability of the wave in the domain at any point in time. This behavior is indicative of the presence of long period tsunami waves, which frequently occur in the interaction of tsunamis with island groups in the absence of a continental shelf.

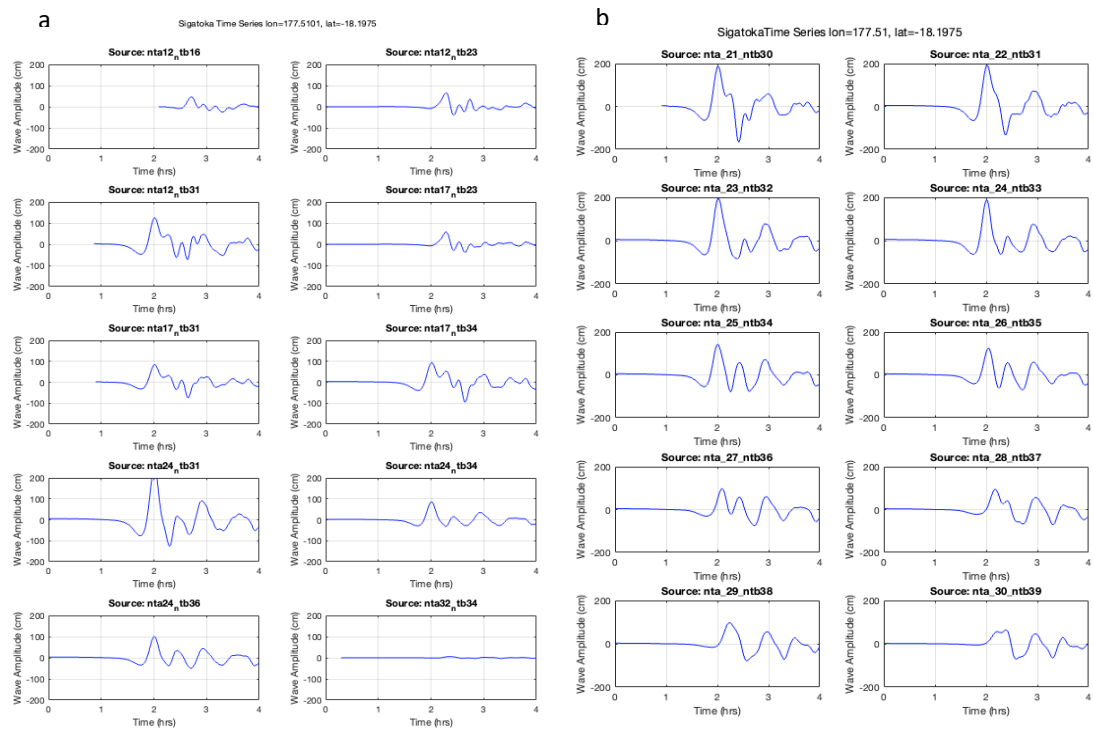

*Figure 8: Representative time series of the sensitivity test for the community of Sigatoka from the ensemble of expert sources along the Tonga-Kermadec trench (panel a) and from the systematic selection along the same trench (panel b). Showing the worst-case scenarios (Tonga-Kermadec: nta_24_ntb31 and Tonga-Kermadec Experts: nta23-ntb32).*

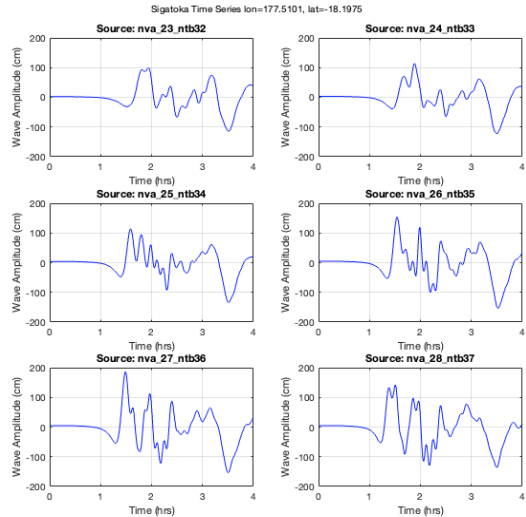

*Figure 9: Representative time series of the sensitivity test for the community of Sigatoka from the New Hebrides systematic selection. Showing the worst-case scenario (nva_27_nvb36).*

In order to validate the sensitivity simulations performed on the lower resolution optimized grids, a comparison of the timeseries obtained when the selected wort-case sources were run using the high-resolution grids revealed excellent agreement between the low-resolution and high-resolution results, further validating the approach of running the optimized grids for worst-case scenario identification. Figure 10 shows the time series comparison for Sigatoka for the three selected scenarios for which both the low- and high- resolution models were run. Similar agreement is found in the Cuvu time series.

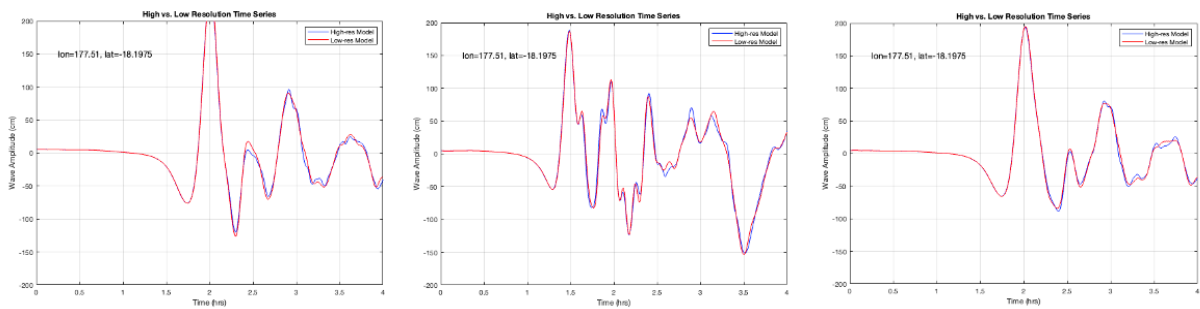

*Figure 10: Time series comparison between the low-resolution model (red) and high-resolution model (blue) for each of the three worst-case scenarios derived from the seismic experts ensemble (left), New Hebrides Trench (center) and Tonga-Kermadec (right) for the community of Sigatoka.*

Overall, one final worst-case scenario candidate was selected in this manner from the New Hebrides trench (nva27_nvb36), two from the Tonga-Kermadec trench one following the systematic methodology described (ntab23_ntb32), and one selected from the set of sources identified by the seismic experts report (ntab24_ntb31). An additional source candidate, identified in a recent probabilistic tsunami hazard assessment study conducted by NCTR staff for the city of Suva following the methodology prescribed by the American Society of Civil Engineering (ASCE) (ASCE Report) was added to the final set. Tables 3 and 4 below present details of the selected sources and Figure 11 shows initial ocean surface deformation for each source.

*Table 3: Worst case scenario source definitions in terms of NCTR's propagation database of unit sources.*

| Source Codename and Subduction Zone Origin | Unit Sources | Peak Amplitude at Sigatoka probe point |
|---|---|---|
| Tonga-Kermadec | 25*nt23b+25*nt23a+25*nt24b+25*nt24a+25*nt25b+25*nt25a+25*nt26b+25*nt26a+25*nt27b+25*nt27a+25*nt28b+25*nt28a+25*nt29b+25*nt29a+25*nt30b+25*nt30a+25*nt31b+25*nt31a+25*nt32b+25*nt32a | 195.2 cm |
| New Hebrides | 25*nv27b+25*nv27a+25*nv28b+25*nv28a+25*nv29b+25*nv29a+25*nv30b+25*nv30a+25*nv31b+25*nv31a+25*nv32b+25*nv32a+25*nv33b+25*nv33a+25*nv34b+25*nv34a+25*nv35b+25*nv35a+25*nv36b+25*nv36a | 187.1 cm |
| Tonga Kermadec (Experts Report) | 31.25*nt24b+31.25*nt24a+31.25*nt25b+31.25*nt25a+31.25*nt26b+31.25*nt26a+31.25*nt27b+31.25*nt27a+31.25*nt28b+31.25*nt28a+31.25*nt29b+31.25*nt29a+31.25*nt30b+31.25*nt30a+31.25*nt31b+31.25*nt31a | 244.3 cm |

*Table 4: Fault plane parameters used in the definition of the probabilistic ASCE source.*

| ASCE Source | | | | | | | | | |
|---|---|---|---|---|---|---|---|---|---|
| Fault | Lon (deg) | Lat (deg) | Length (km) | Width (km) | Dip (deg) | Rake (deg) | Strike (deg) | Depth (km) | Slip (m) |
| 1 | 185.385792 | -21.01958 | 85.7 | 51.4 | 17.15 | 91.79 | 203.8 | 4.56 | 12.13 |
| 2 | 185.052662 | -21.70949 | 84.2 | 53.3 | 16.62 | 93.83 | 205.5 | 4.03 | 31.5 |
| 3 | 184.694675 | -22.40541 | 87.2 | 52.8 | 16.24 | 89.21 | 205.4 | 4.2 | 31.5 |
| 4 | 184.286850 | -23.06148 | 86.2 | 57.8 | 17.13 | 73.35 | 218.6 | 7.8 | 31.5 |
| 5 | 184.048461 | -23.71295 | 81.6 | 61.1 | 17.86 | 80.31 | 242.6 | 15.59 | 31.5 |

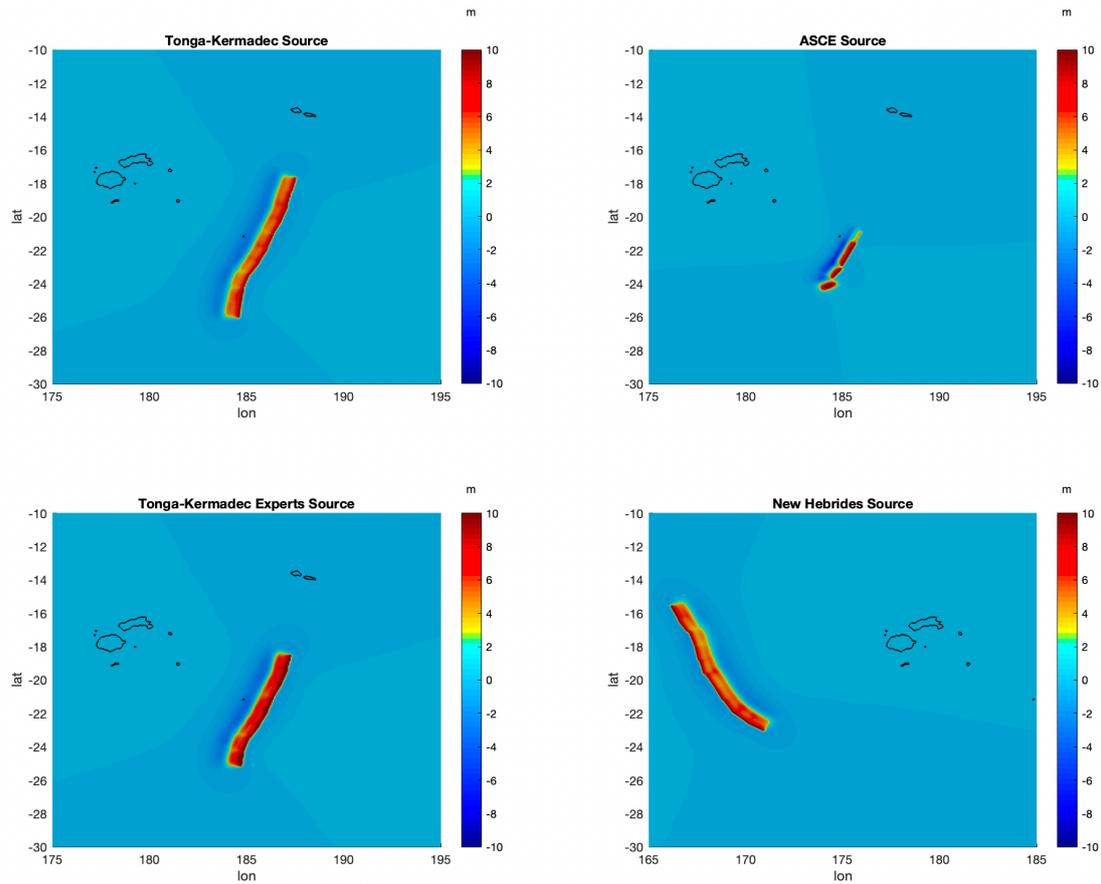

*Figure 11: Initial ocean-surface deformations of the four selected worst-case scenarios.*

For this hazard assessment analysis, no further selections were made. The final inundation map was built as a composite of results generated by each of these source candidates.

# 4. Results

The MOST code was run on the high-resolution set of grids (10-meter) for the communities of Sigatoka and Cuvu, for each of the 4 worst-case sources with initial deformations shown in Figure 11. Output from the code provides values of wave amplitude and velocity components at each grid node as a function of time. Other quantities of interest such as travel time, wave amplitude, flow depth, current speed, and mass and momentum flux can be derived from the output using postprocessing software. Of particular interest for the development of tsunami evacuation maps are the maximum wave amplitude and max flow depth. Maximum current speed is important in assessing tsunami impact on harbor facilities and for coastal navigation. Tsunami generated currents can be responsible for substantial damage in harbor infrastructure even in the case of events with little overland inundation.

For all inundation simulations in the study, wave amplitude and velocity components at every grid node were output every 45 seconds of simulated time. However, some important parameters such as maximum wave amplitude and velocities are internally tracked by the code at every time step (time step=0.24 seconds) for maximum temporal resolution and output as a separate file at the completion of the simulation. This maximum amplitude file was used to derive the maximum flow depth values.

To compute the final inundation area, a composite map reflecting the maximum wave amplitude (alternatively flow depth) from any of the four sources simulated was constructed. This composite maximum map is, therefore, a maximum of maxima, assigning to any grid node the largest value of wave amplitude (or flow depth) recorded for any of the four worst-case sources and is typically used as the basis for evacuation mapping and emergency planning.

All four sources simulated in the study originate in the near field of Fiji, travel time calculations reveal a very similar arrival time for all of them with the first waves arriving approximately 1.5 hrs after the event at the communities of Sigatoka and Cuvu as evidenced in the time series of Figure 10 for the case of Cuvu. However, arrival times may range from 1 to 2 hours for some of the less hazardous sources located along the Tonga-Kermadec or New Hebrides trench also evident from Figures 8 and 9.

## 4.1 Tsunami Propagation

Of the four worst-case scenarios simulated at high-resolution, the deep water solution for three of them, namely, the two obtained from the sensitivity analysis along the Tonga-Kermadec and New Hebrides margins and the one selected from the ensemble prescribed by the local seismic experts report, was obtained from NCTR's propagation database by linearly combining and scaling the individual solutions of appropriate unit sources to provide boundary and initial values to the set of inundation grids centered in the communities of interest. Retrieval from NCTR's database was not possible in the case, of the ASCE study source as this one is not defined in terms of NCTR's standard unit sources, therefore deep-water propagation for this source was computed independently.

Figure 12 shows the distribution of the wave's maximum amplitude in deep water for each of the sources selected. The directivity of the source energy can easily be assessed from the image. Of all four sources, the two located along the Tonga-Kermadec margin seem to have the largest impact on the Fiji Islands. However, Figures 8 and 9 reveal that the largest amplitude waves on the Coral Coast are generated by the Tonga-Kermadec Experts source. Suggesting that this source is most likely the main contributor to the composite inundation maps for Sigatoka and Cuvu.

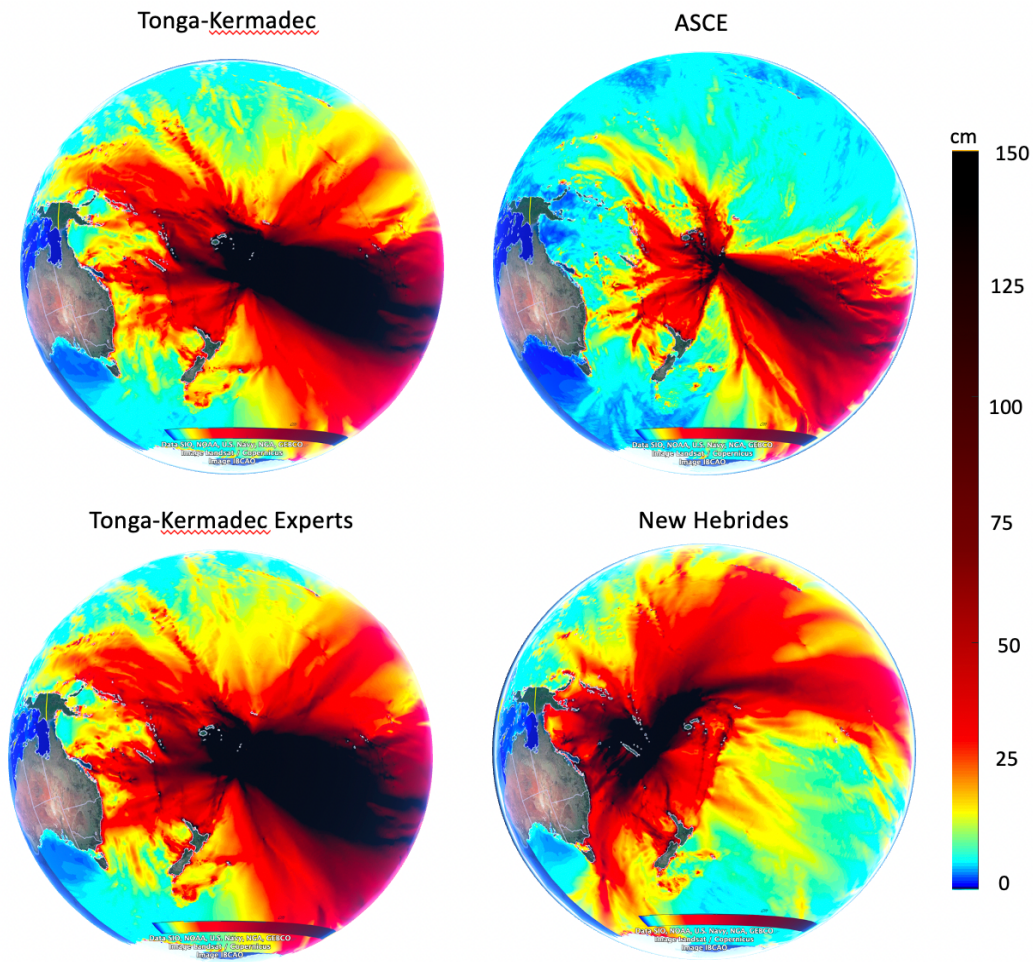

*Figure 12: Tsunami maximum wave amplitude distribution in deep water for Tonga-Kermadec (upper-left), Tonga-Kermadec Experts (lower left), ASCE (upper right) and New Hebrides (lower right) sources investigated. The two sources from Tonga-Kermadec appear to have similar impact on the Fiji Islands, however the time series of Figure 8 reveal that the Tonga-Kermadec Experts source generates the largest waves on the Coral Coast.*

## 4.1 Tsunami Inundation

The MOST model computes tsunami inundation by calculating the solution of the Shallow Water Wave equations on a set of three nested grids centered around the coastal area of interest as described in Section 3.3. Typically, tsunami waves reach the inundation grids through the boundaries of the outer grid (A-grid) defining a classical boundary value problem, except in the case of very near-field sources when co-seismic deformation of the generating source reaches the geographic extent of the A-, B- or C-grids. In such case, the problem becomes a combination of initial value and boundary value problem.

In the particular case of the 4 sources under investigation in this study, all four of them are distant enough from the Fiji Islands that the bulk of the deformation takes place outside of the extent of the A-grid, therefore, the deep-water propagation solution from each of the sources provided boundary conditions for all four sources.

As tsunami waves transition from the deep ocean to more shallow waters, they experience a reduction in propagation speed that results in shortening of their wavelength (and increase in their amplitude). In order to approximately maintain a constant number of grid nodes per wavelength, which is a requirement of the numerical method, the deep-water solution from NCTR's database is transferred to the higher resolution A-grid by interpolation along its boundaries. As the shoaling process continues, tsunami waves are further transferred to increasingly higher resolution grids (B- and C-grids). Finally, along the wet/dry boundary of the highest resolution grid (C-grid) run-up calculations are performed to compute coastal inundation.

Given the modeling methodology used in the study, in which a deep-water propagation solution is constructed by the linear combination of precomputed unit source solutions, it is important to ensure that the deep-water propagation solution is transferred to the high-resolution nested grid set prior to the development of non-linear effects, i.e. before the waves approach shallow waters and the linear assumption breaks down. If any non-linear effects were to develop in the deep-water regions where the solution is being linearly combined, this would invalidate such approach. Once the wave is being computed in the nested set of grids, non-linear calculations are performed and non-linear effects of tsunami waves in shallow waters are adequately captured by the code. This requirement is easily met when modeling tsunami impact on islands as the absence of continental shelf ensures ocean depths grow rapidly away from the shore, effectively reducing the need for the outer and intermediate grids (A- and B- grids) to extend far offshore.

For the purpose of ensuring the complex dynamics of tsunami waves between neighboring islands are properly captured in our simulations, the A- and B-grids were designed so that the A-grid included both neighboring islands of Viti Levu and Vanua Levu and the B-grid included the entirety of Viti Levu island. The precise extents and resolutions of the numerical grids have already been presented in Section 3.3

Figures 13 through 15 below show maximum computed tsunami amplitudes, flow depths and current speed for each of the four selected sources (Tonga-Kermadec, New Hebrides, ASCE and Tonga-Kermadec Experts, for the community of Sigatoka and surrounding areas.

Maximum tsunami amplitudes represented in the figures correspond to the maximum wave elevation reached at any point on the map at any time during the simulation of the corresponding source, since the wave does not peak at all points in the domain simultaneously, not all points reached their maximum elevation at the same instant in time, therefore tsunami maximum amplitude images shown in the figures do not represent a snapshot in time of the simulation. However, peak amplitudes do tend to be associated with the largest wave in the wave-train as it propagates from deep into more shallow waters and then runs up. The same explanation applies to the other magnitudes being represented: tsunami flow depth and current speed.

It is also important to clarify the distinction between tsunami maximum amplitude and flow depth. As mentioned, earlier maximum tsunami amplitude represents the maximum elevation of the wave at any point measured from the vertical datum of the study, i.e., Mean High Water (located at the shoreline). Maximum flow depth also represents the maximum elevation of the wave, but only on land (over flooded areas) and measured vertically from the local topographic elevation to the water surface, therefore flow depth is equivalent to the maximum depth of water that a person standing at a particular location would experience measured from his/her feet to the water surface.

Lastly, current speed is the maximum strength of the current generated by tsunami waves at any point on the study region. Current speed can be a critical parameter to monitor during a tsunami event, specially, in those cases in which tsunami waves arrive at low tide and do not cause major inundation. Tsunami current speeds are known to have caused major damage to harbor facilities in cases with minimal overland inundation, for instance in coastlines with a large tidal range, when substantial tsunami waves arrive at low tide.

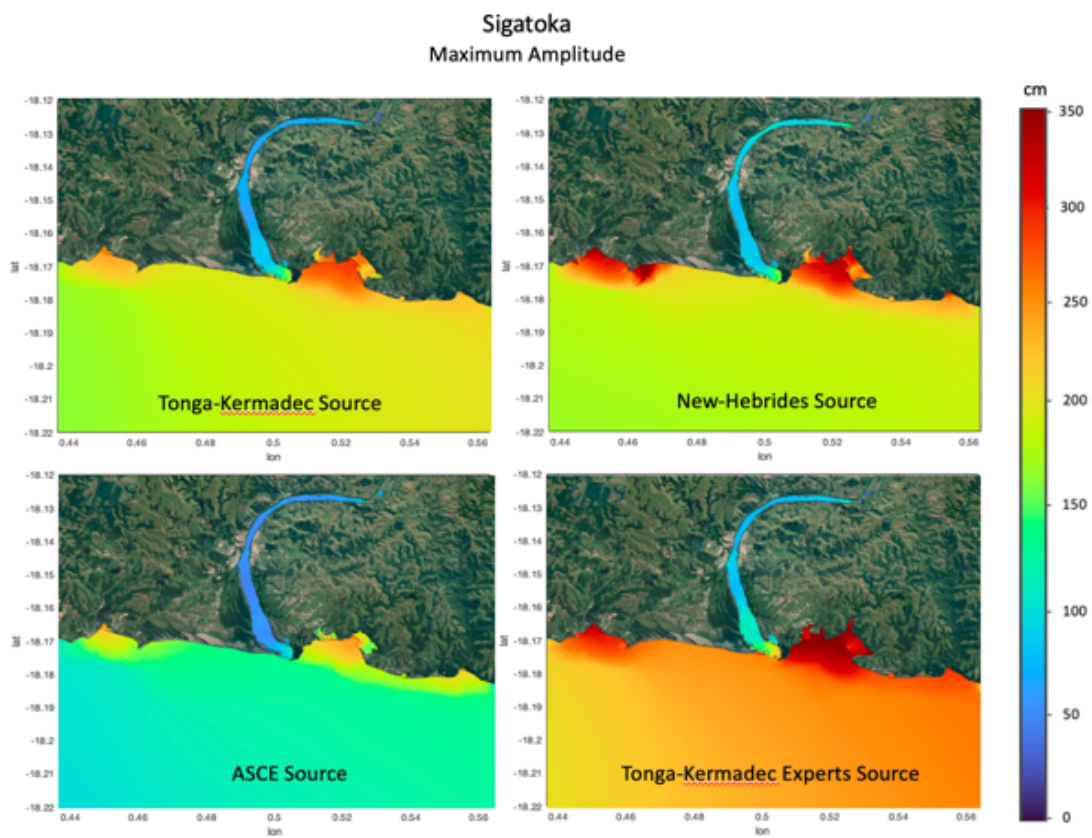

*Figure 13: Tsunami Maximum Amplitude Plots for the community of Sigatoka, overlayed over Google Earth maps, for each of the worst-case scenarios investigated. The colors represent the maximum wave elevation computed at that location throughout the simulation.*

Based on the maximum amplitude plots of Figure 13, the most hazardous source of the four selected seems to be the Tonga-Kermadec source designed by in the IOC Workshop of local experts (2008) as already suggested by the results of Figure 12 and the time series of Figure 8. Consequently, this scenario is seen to dominate the composite maximum amplitude plot.

Tsunami maximum flow depths plots shown on Figure 14 are consistent with the maximum amplitude graphics of the Figure 13 and show most of the flow depth being generated by the Tonga-Kermadec Experts source and maximum flow depths well in excess of two meters in some areas.

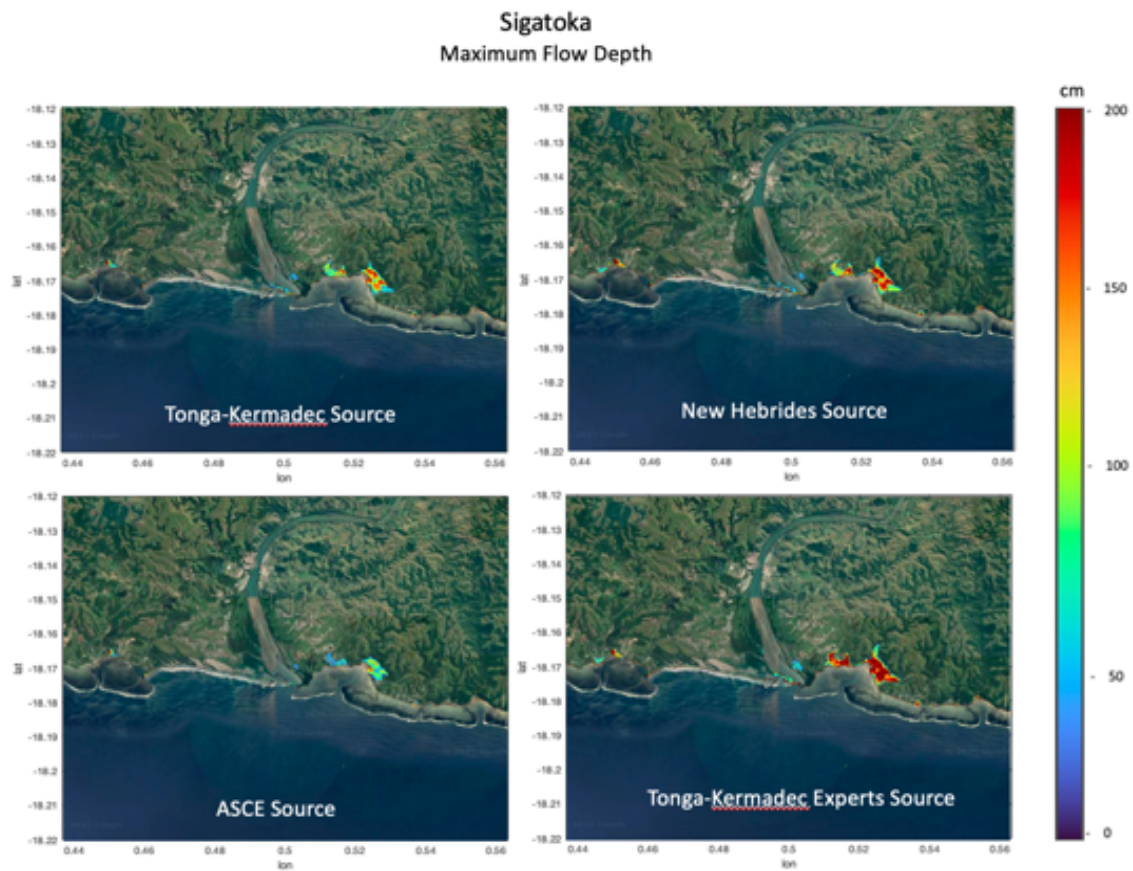

*Figure 14: Tsunami Maximum Flow Depth plots for the community of Sigatoka, overlayed over Google Earth maps, for each of the worst-case scenarios investigated. The colors represent the maximum wave elevation computed at that location throughout the simulation measured from the ground to the water surface.*

Figure 15 shows maximum values of the water current speed generated in Sigatoka and surrounding areas by each of the four different scenarios selected. In all cases current speeds are shown to exceed 4 m/s over land and in the coral reef. It is interesting to note that although the New Hebrides source is not the one that generates the highest wave amplitudes, it is the one generating the strongest currents, particularly in the coral reef area. This is probably due to the generation of eddies and vortical structures in the velocity field within the coral reef which typically exhibit very high spiraling velocities, and are quite hard to predict in a deterministic way.

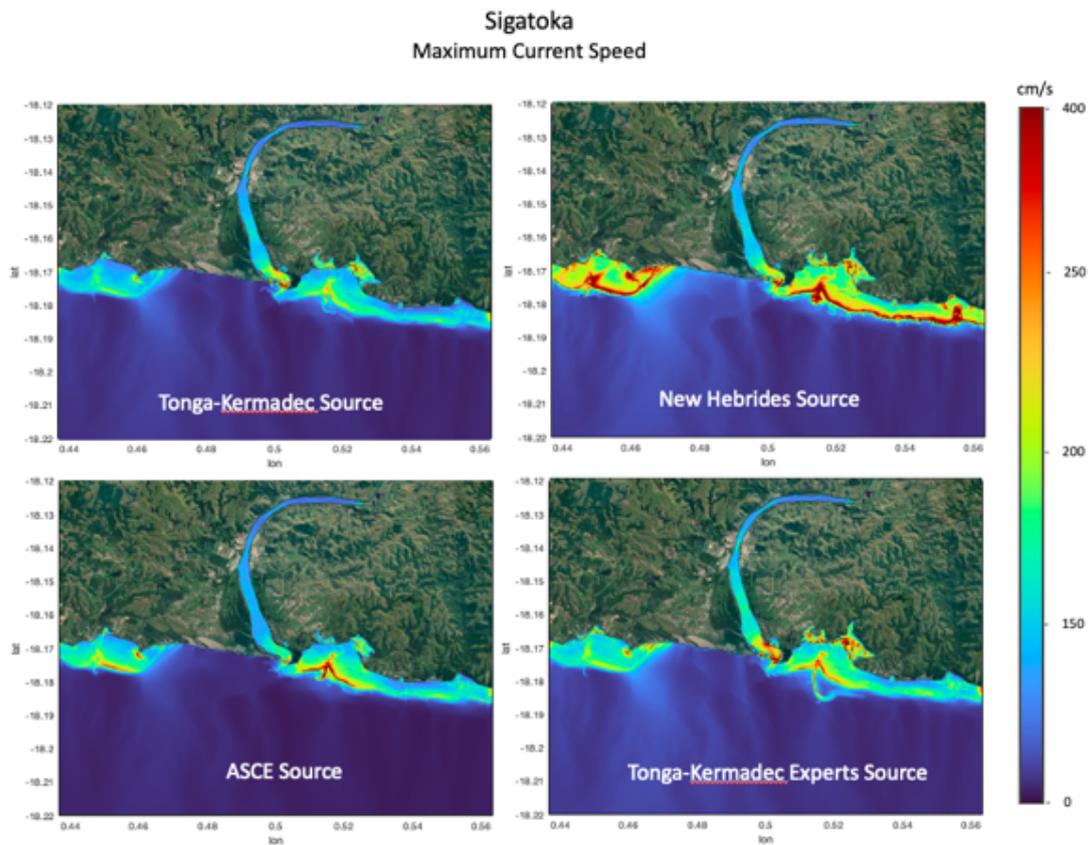

*Figure 15: Maximum computed tsunami Current Speed for each of the four selected sources in Sigatoka and surrounding areas. The colors represent the maximum current speed computed at that location throughout the simulation. Based on the images, the New Hebrides source becomes dominant in the composite maximum current speed map.*

The final composite Maximum Tsunami Amplitude, Flow Depth and Current Speed map for Sigatoka and surrounding areas is represented in Figures 16 through 18. For convenience, the lower panel of Figure 16 shows the maximum tsunami wave amplitude along the coastline of Sigatoka. As expected from the data shown in panel A, Maximum Tsunami Amplitude is dominated by the Tonga-Kermadec Experts scenario and looks very similar to the results shown on the lower right panel of Figure 13, while the Maximum Current Speed map is dominated by the New Hebrides source and looks similar to the upper right panel of Figure 15.

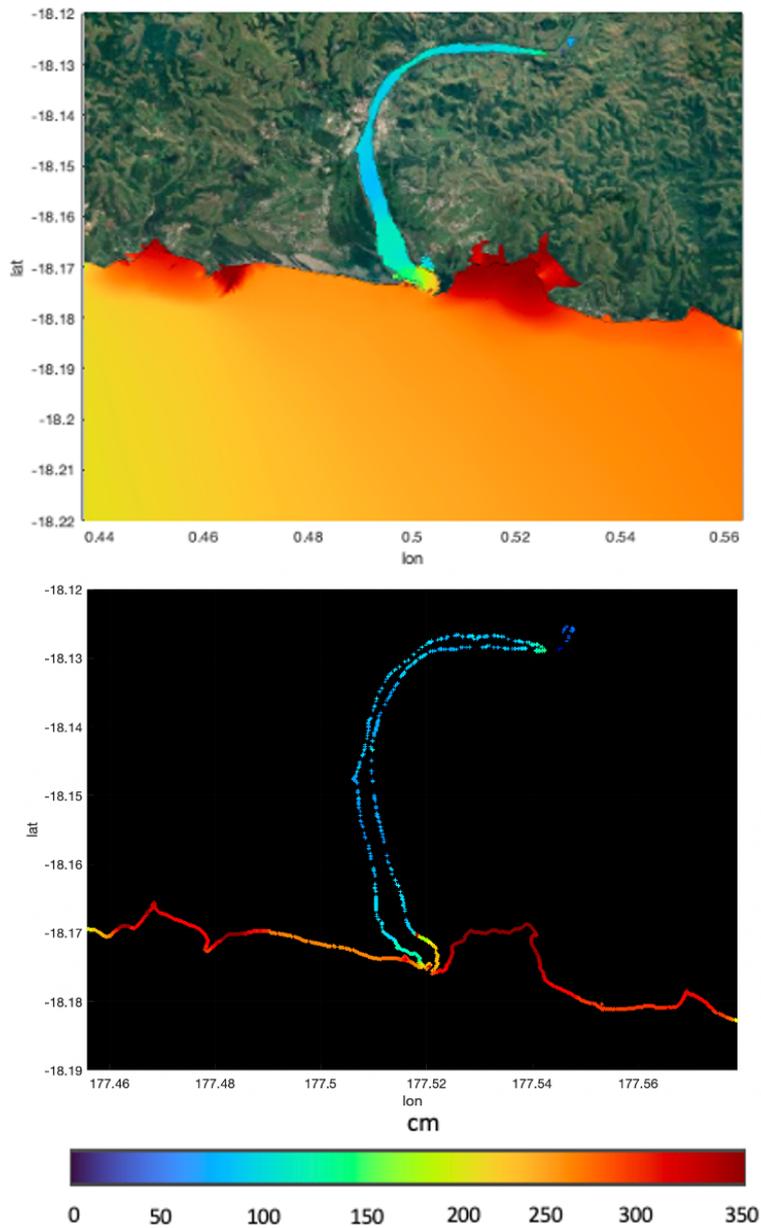

*Figure 16: Composite Maximum Tsunami Wave Amplitude in Sigatoka. Results from the study show that a small amount of energy will penetrate and propagate upstream the Sigatoka River causing very little inundation along its shores and in the main population center of Sigatoka itself. Lower Panel shows maximum wave amplitudes along the coastline.*

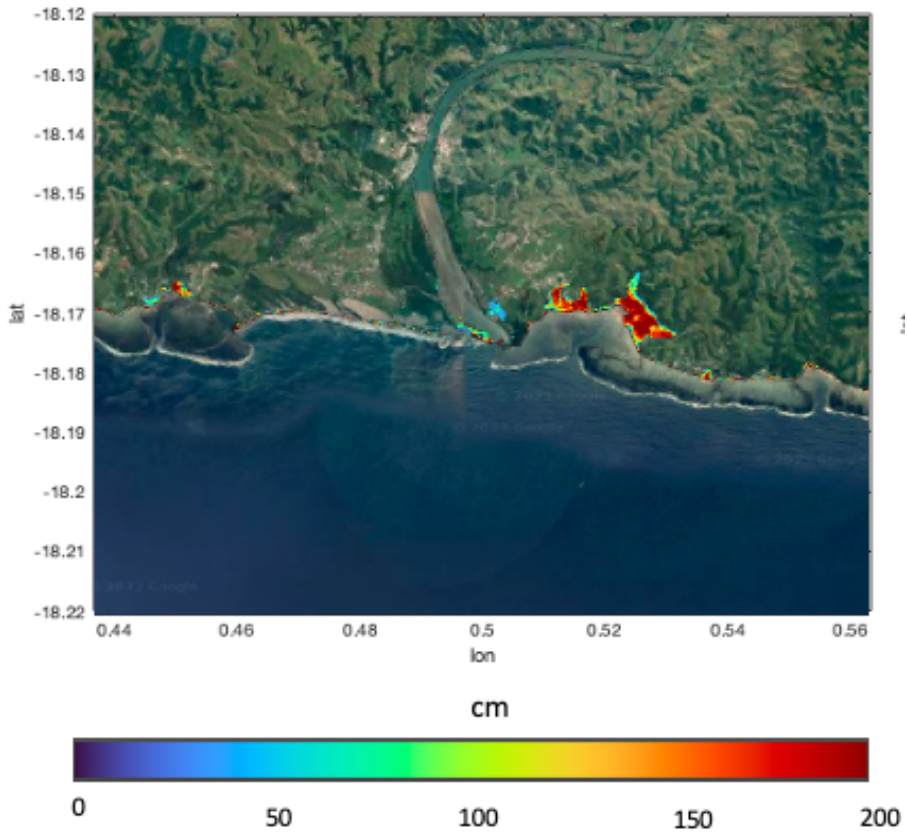

*Figure 17: The Composite Flow Depth map for Sigatoka and its surroundings is shown in the image. Consistent with the Maximum Amplitude map, most of the inundation takes place along the coastline to the east of the river mouth and along the Sigatoka River banks immediately upstream of the river mouth.*

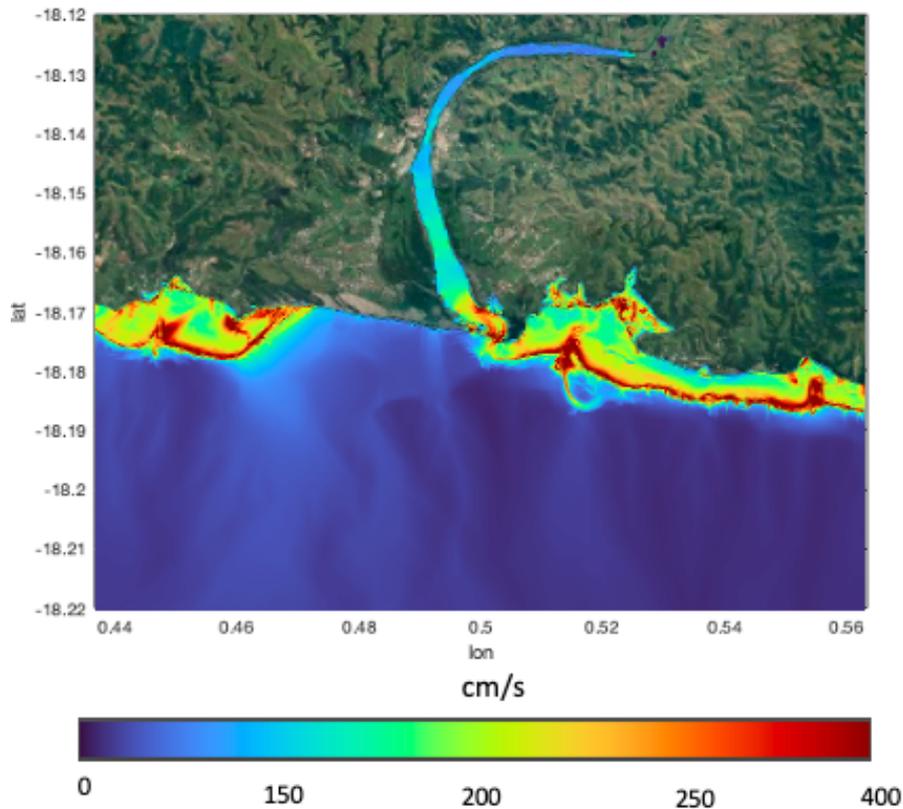

*Figure 18: The Composite Maximum Current Speed map for Sigatoka and its surroundings is shown in the image. Consistent with the Maximum Current Speed map of Figure 15. The strongest currents develop in and around the coral reef, very likely due to the generation of vortical structures in the velocity field.*

The results of the study show a limited amount of wave energy penetrating the Sigatoka River and travelling upstream to the town of Sigatoka itself. This small amount of energy is not sufficient to generate significant inundation along the banks of the river except at some locations immediately upstream from the river mouth. Tsunami modeling upriver is undoubtedly affected by the lack of riverine bathymetric information in the provided DEM as well as, perhaps, a somewhat inaccurate representation of the river mouth entrance as observed in discrepancies between the DEM and Google Earth imagery for this area. This should be a priority area for DEM improvement in future studies.

Figure 19 through 21 show maximum computed tsunami amplitudes, flow depths and current speed for each of the four selected sources (Tonga-Kermadec, New Hebrides, ASCE and Tonga-Kermadec Experts) for the community of Cuvu and surrounding areas immediately to the west of Sigatoka.

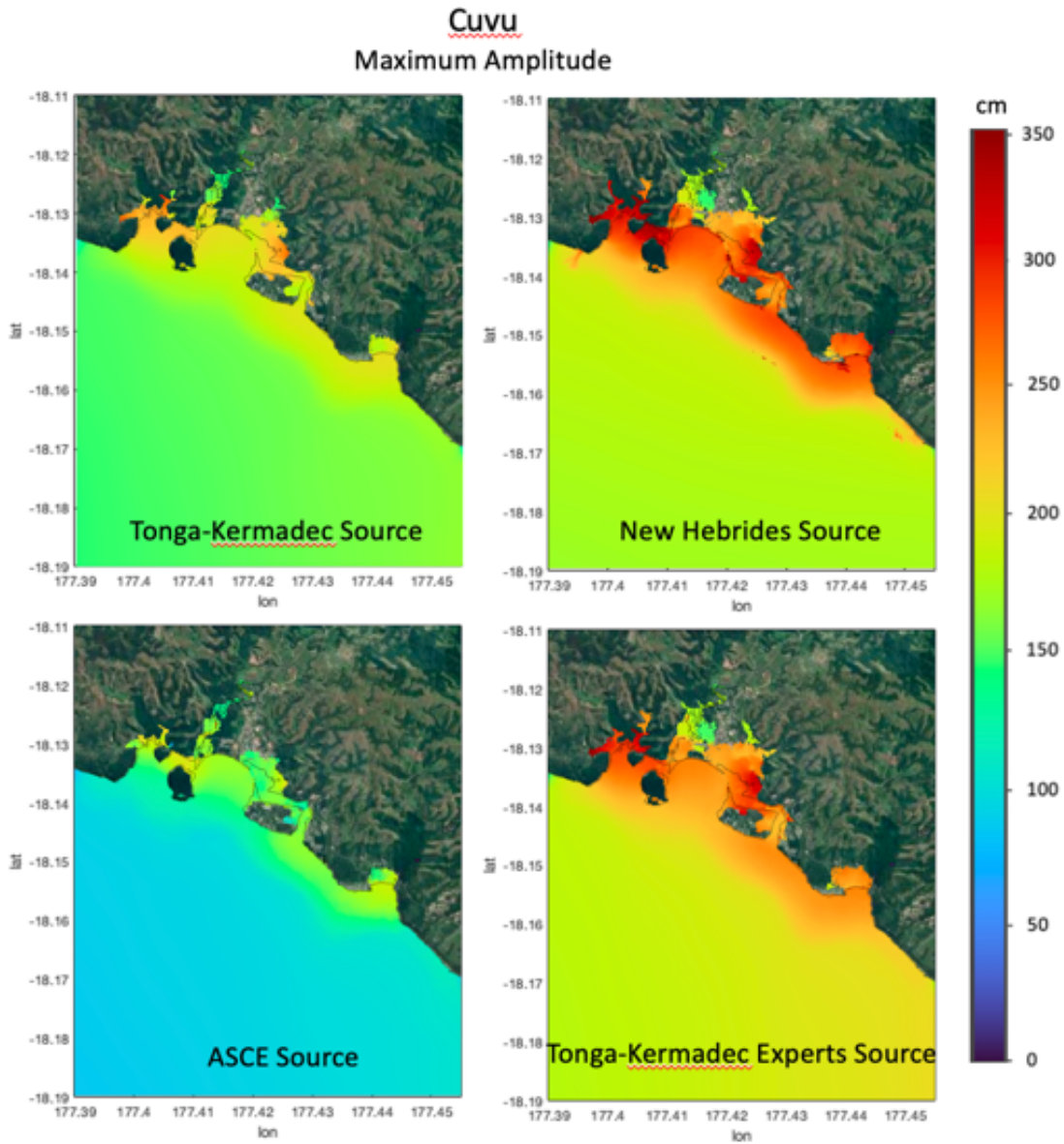

*Figure 19: Tsunami Maximum Amplitude Plots for the community of Cuvu, overlayed over Google Earth maps, for each of the worst-case scenarios investigated. The colors represent the maximum wave elevation computed at that location throughout the simulation.*

In the case of Cuvu, both the source scenario from the New Hebrides trench and the Tonga-Kermadec selected by the local experts are observed to be quite similar and dominant, in terms of the maximum tsunami amplitudes they generate along the coastline of Cuvu.

Figure 20 shows the maximum computed Flow Depth for each of the four selected sources in and around Cuvu. Consistent with Figure 19, maximum flow depth is dominated by both the New Hebrides and the Tonga-Kermadec Experts source with values in excess of 3.5 meters present throughout the coastline in any of the four scenarios, but most prominent in the two dominant ones.

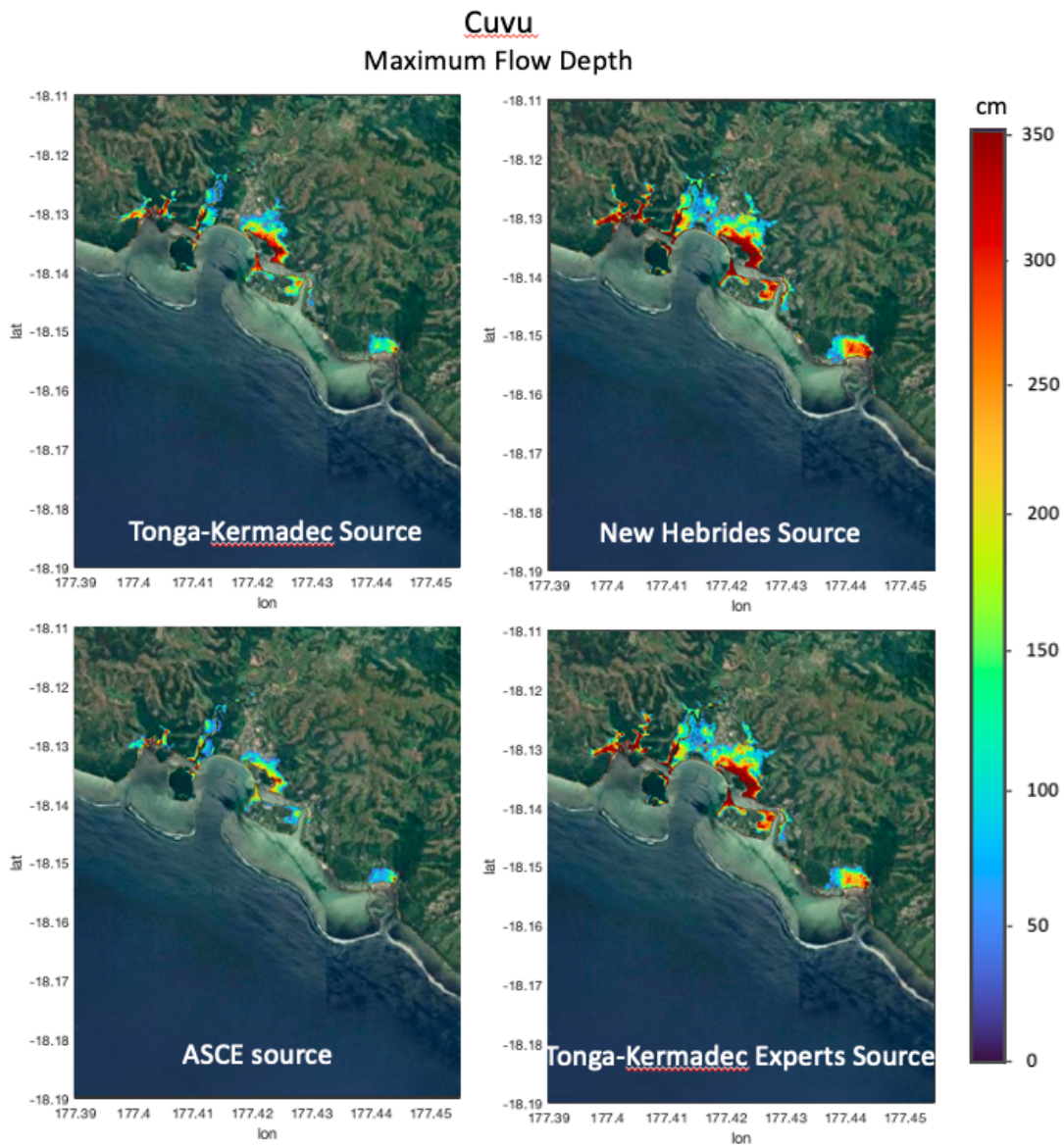

*Figure 20: Tsunami Maximum Flow Depth plots for the community of Cuvu, overlayed over Google Earth maps, for each of the worst-case scenarios investigated. The colors represent the maximum wave elevation computed at that location throughout the simulation measured from the ground to the water surface.*

.

The New Hebrides source scenario continues to dominate in terms of the generation of strong currents in the area around Cuvu with large areas of current speed higher than 4 m/s along the coast, but specially within the coral reef, as seen in Figure 21.

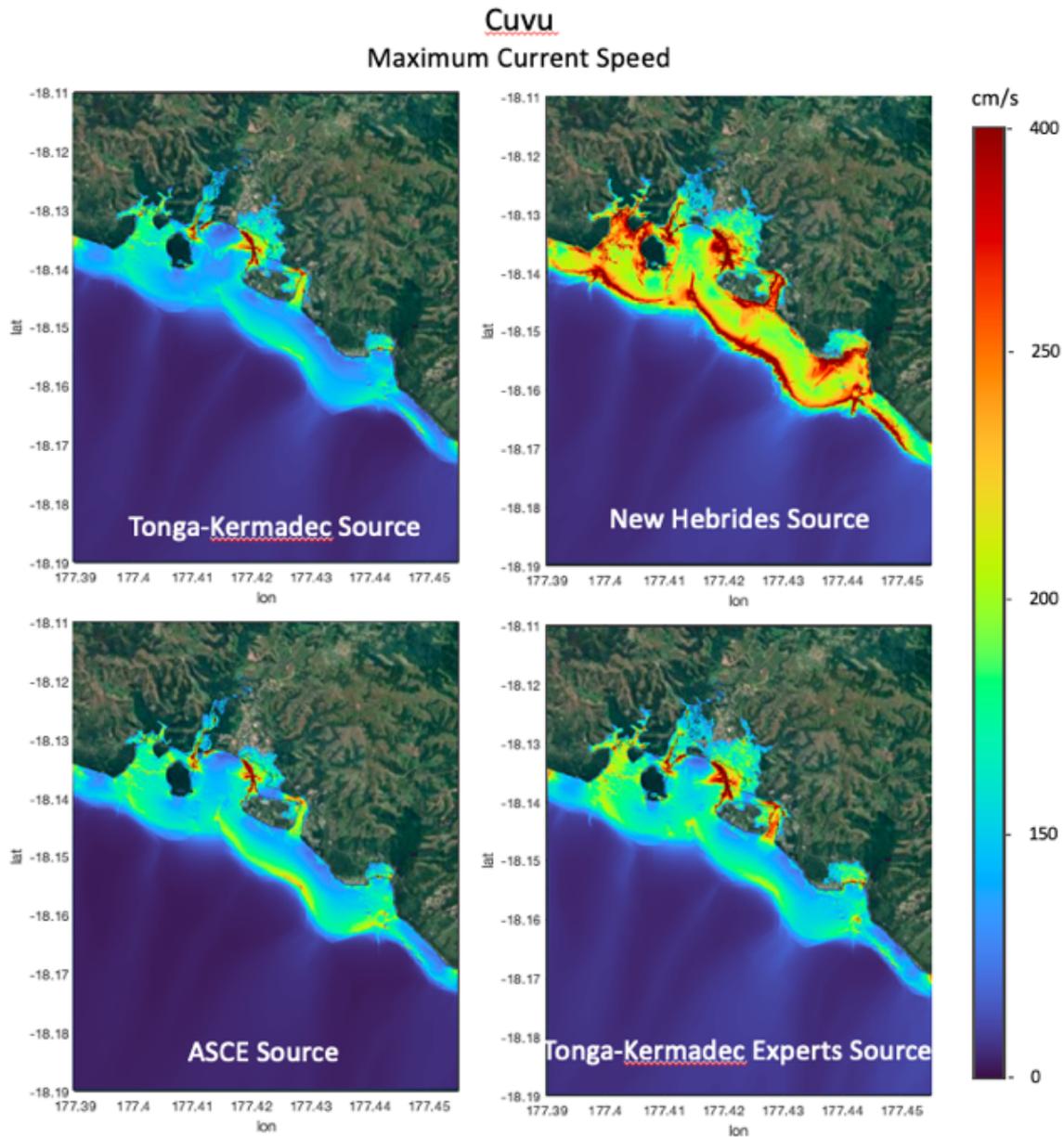

*Figure 21: Maximum computed tsunami Current Speed for each of the four selected sources in Cuvu and surrounding areas. The colors represent the maximum current speed computed at that location throughout the simulation. Based on the images, the New Hebrides source becomes dominant in the composite maximum current speed map.*

The final composite maps of Maximum Tsunami Amplitude, Flow Depth and Current Speed for Cuvu and surrounding areas are represented in Figures 22 through 24. For convenience, Figure 22 (lower panel) shows the maximum tsunami wave amplitude along the coastline of Cuvu. As expected from the data shown in Figure 19, Maximum Tsunami Amplitude is dominated by both the New Hebrides scenario and the Tonga-Kermadec Experts scenario, while the Maximum Current Speed map is dominated mainly by the New Hebrides source.

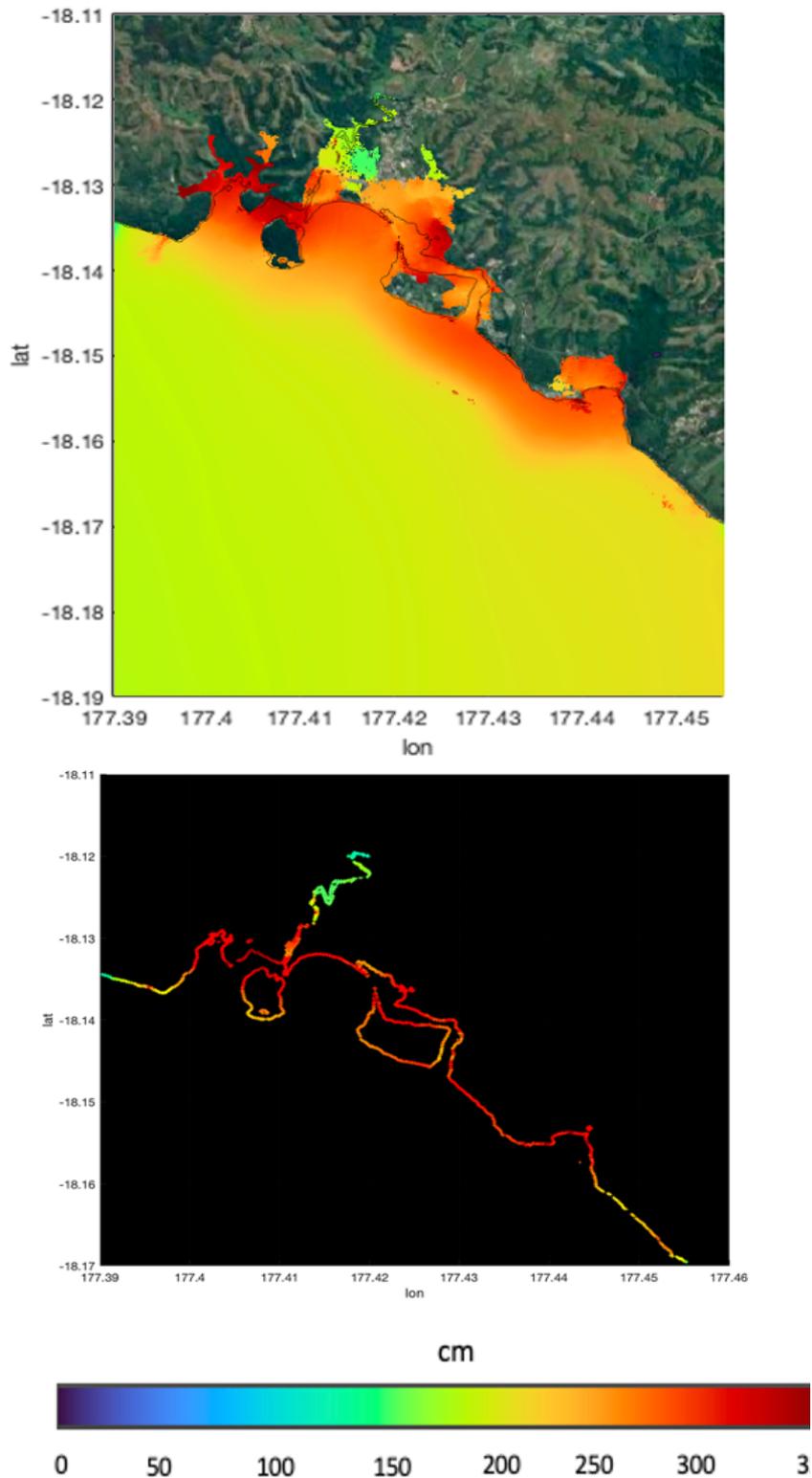

*Figure 22: Composite Maximum Tsunami Wave Amplitude in Cuvu. Results from the show tsunami waves in excess 3 meter in amplitude hitting the coastline of Cuvu and Voua and inundated large extents of low-lying areas populated areas. Lower Panel shows maximum wave amplitudes along the coastline.*

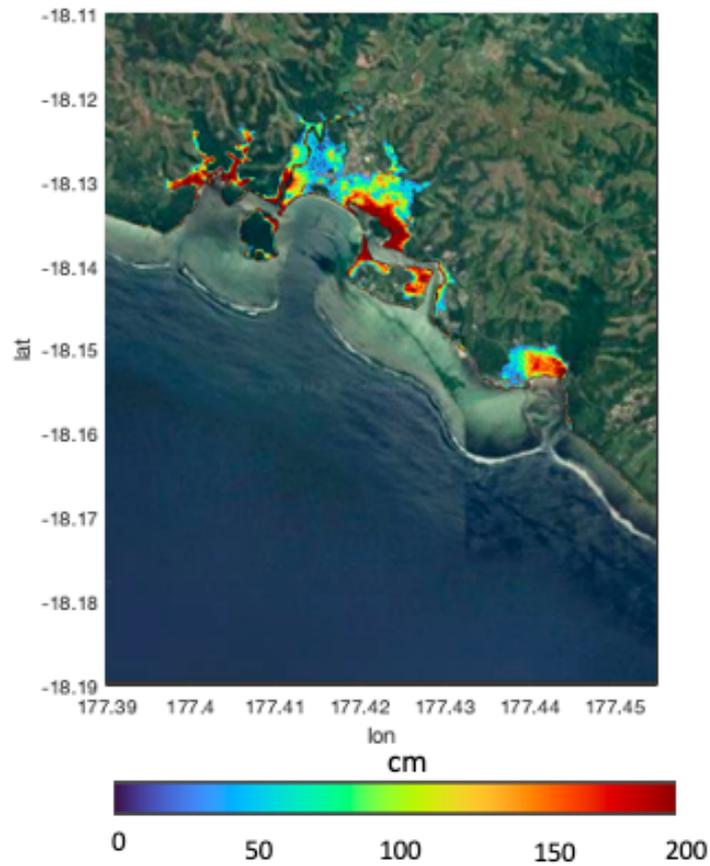

*Figure 23: The Composite Flow Depth map for Sigatoka and its surroundings is shown in the image. Consistent with the Maximum Amplitude map, extensive, low-lying areas along the coast are flooded with flow depths up to 2 meters.*

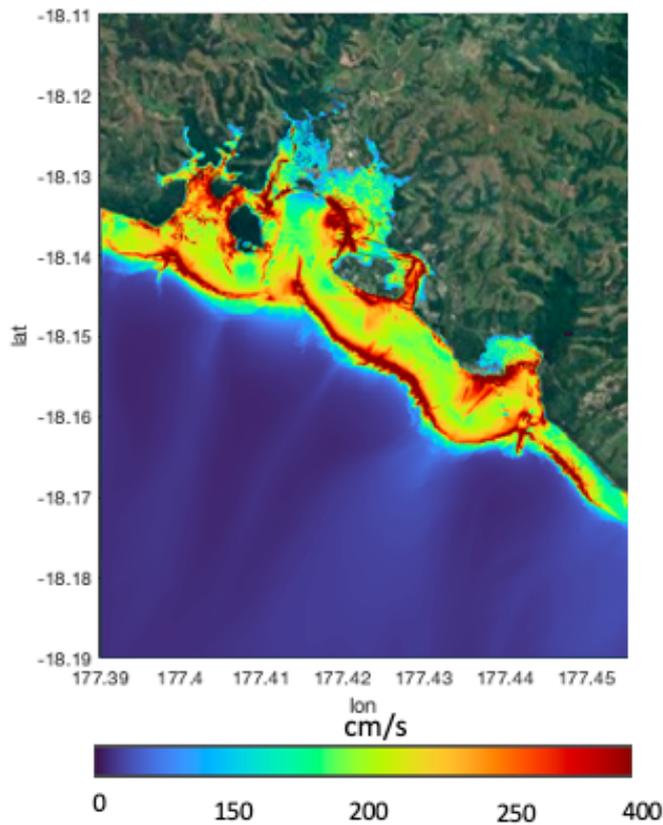

*Figure 24: The Composite Max. Current Speed map for Cuvu and its surroundings is shown in the image for each individual scenario. Consistent with the Maximum Current Speed map of Figure 21, the strongest currents develop in and around the coral reef, very likely due to the generation of vortical structures by the New Hebrides source scenario.*

## 4.2 Temporal Analysis

In addition to the Sigatoka time series presented in the sensitivity analysis of Section 3.4.2, we present here the time series associated with both tsunami wave amplitude and current speed for each of the four selected sources and for the two communities of Sigatoka and Cuvu. The Sigatoka time series have been taken at the mouth of the Sigatoka River (lon=177.5168°, lat=-18.1777°) at a point representative of shallow water values prior to tsunami propagation up the Sigatoka River. In the case of Cuvu, the sample point was taken at the center of Cuvu Harbor (lon=177.4161°, lat=-18.1353°) and is representative of the tsunami amplitudes and currents that develop in an area unprotected by the coral reef. Figure 25 shows the two locations were time series were extracted on a google Earth map. Figures 26 for Sigatoka and 27 for Cuvu show all four tsunamis arrive as a leading elevation wave followed by decaying smaller waves. In the cases of Tonga-Kermadec, New Hebrides and Tonga-Kermadec Experts sources, this first leading elevation wave is also the largest in the record. This is however, not the case for the ASCE source, which shows a relatively mild leading elevation wave followed by a pronounced trough and then a second elevation wave which is the highest in the wave train with the rest of wave amplitudes decaying rapidly after this one.

The current speed time series at the same points indicate that currents developed in the New Hebrides source scenario tend to be stronger than in any of the other events, peaking at over 2 m/s at the location where the timeseries has been sampled. This is consistent with the results presented so far and probably indicative of strong vortex formation throughout the coral reefs for tsunamis from this region, but somewhat unintuitive since this scenario does not generate the largest wave amplitudes overall as seen in the images of Figures 13 and 19.

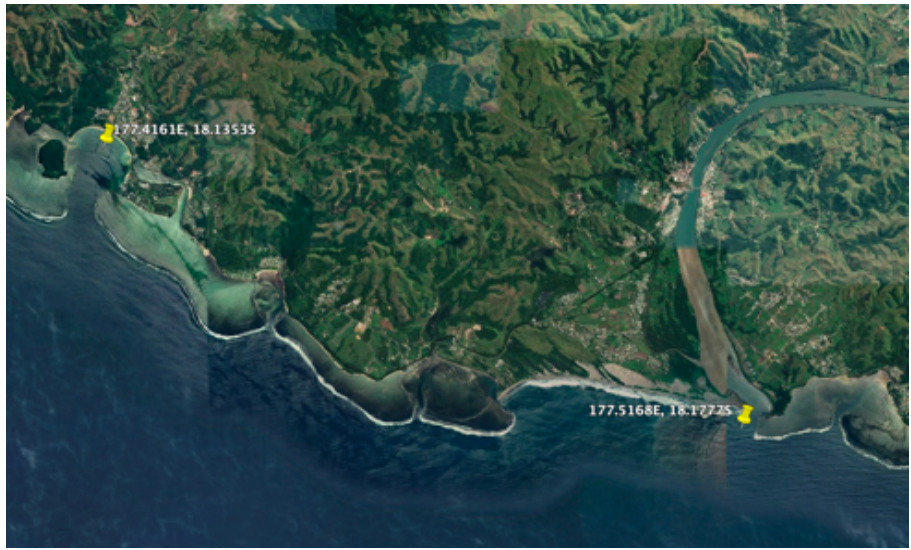

*Figure 25: Precise Time series locations for the entrance of the Sigatoka River and Cuvu Harbor.*

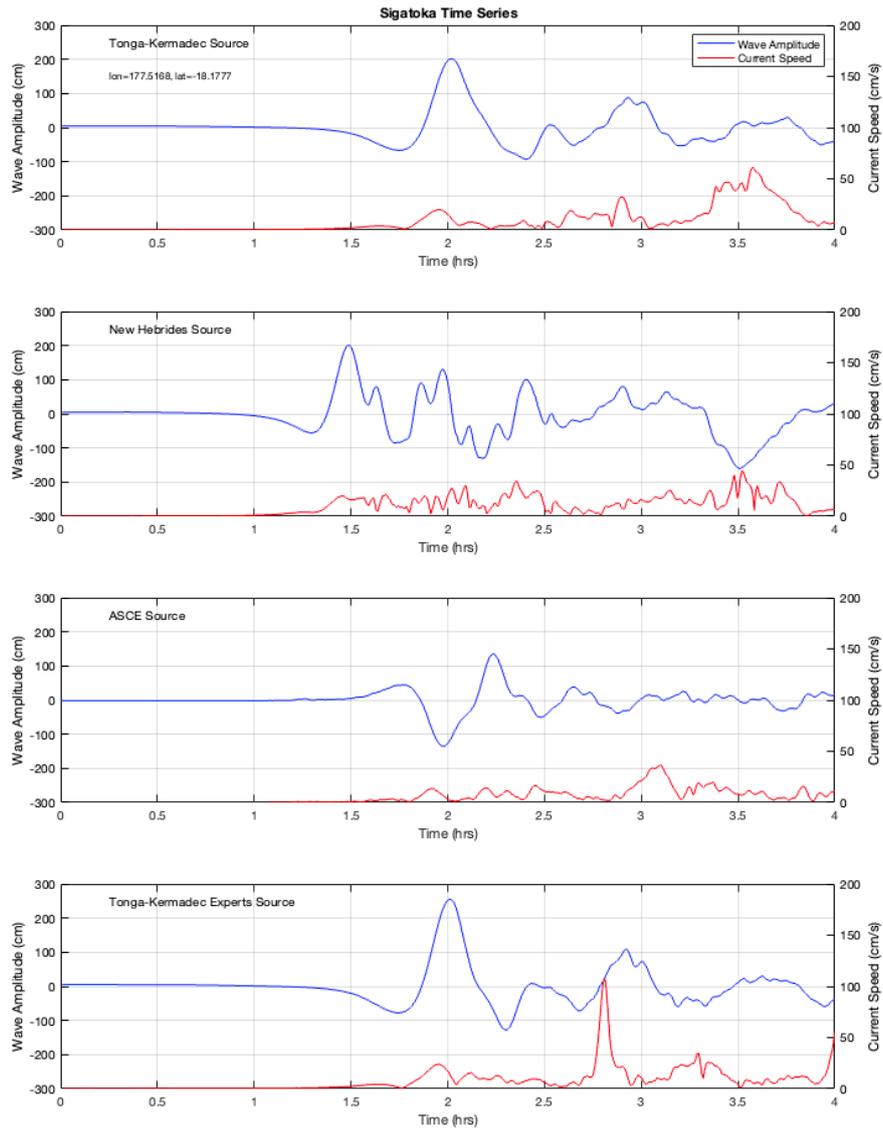

*Figure 26: Times series of tsunami wave amplitude and current speed at the entrance of the Sigatoka River (lon=177.5168, lat=-18.1777).*

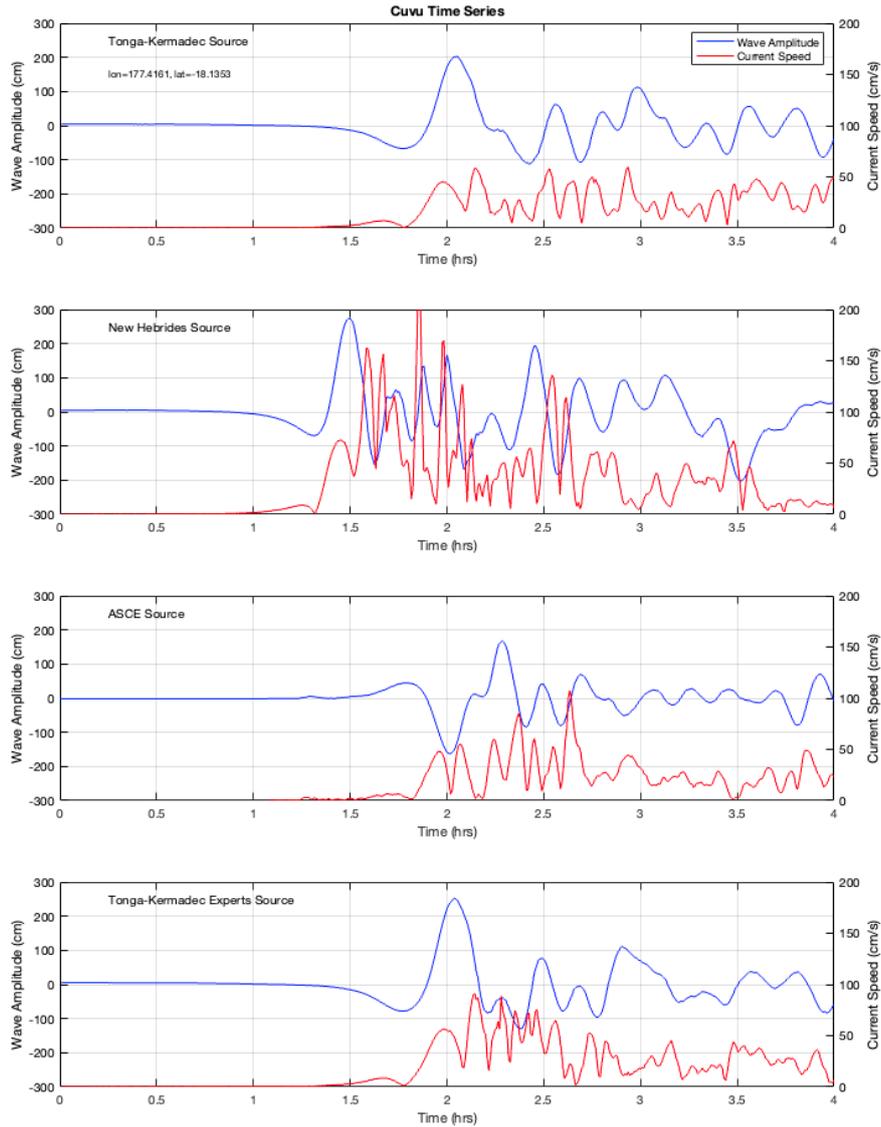

*Figure 27: Times series of tsunami wave amplitude and current speed at Cuvu Harbor (lon=177.4161, lat=-18.1353).*

# Conclusion

The objective of this project is to perform a tsunami hazard assessment study for two communities located along the Coral Coast of Fiji. The Republic of Fiji sits in close proximity to some of the most active regions of the Pacific Ocean "Ring of Fire" and is vulnerable to tsunamis generated in both the Tonga-Kermadec and New Hebrides trench. A combined approach was used for the identification and selection of worst-case scenarios for the locations under investigation here, Sigatoka and Cuvu. One approach was to research the existing literature for historical and potential seismic sources that could affect Fiji. Following this approach two scenarios were selected, one is a scenario identified in a recent study conducted by NCTR for the US Dept. of State for the community of Suva, conducted according to methodology prescribed by the American Society of Civil Engineering, the second one was selected from an ensemble of potential sources identified in a 2018 IOC Report on Tsunami sources form the Tonga-Kermadec trench. Finally, two other scenarios were based on a systematic, sensitivity study, performed using NCTR's Tsunami Propagation Database (Gica et al., 2008) conducted to model tsunami impact along the coastlines of interest. Two scenarios were selected using this approach from a total of 56 discrete earthquake sources located along the New Hebrides and Tonga-Kermadec subduction zones, which lie in close proximity to the Fiji Islands. Arrival times to the communities of interest for waves from these sources range from 1 to 2 hours after the earthquake and are very similar for both study sites.

Modeling results show substantial areas of both Sigatoka and Cuvu being inundated by tsunami waves with waves up to 5 meters along the shoreline with the two most dangerous tsunamis being located one to the west of Fiji in the New Hebrides subduction zone and one in the Tonga region of the Tong-Kermadec subduction zone.

Observed inundation patterns are similar to those seen in most coastlines with low-lying beach areas experiencing the worst inundation. Of particular interest, however, is the effect the presence of the coral reef has on tsunami dynamics. While in general, the reef tends to either dampen or reflect tsunami energy protecting coastal areas immediately onshore of the reef, reef openings definitely pose a risk to those beach areas located immediately onshore. It is possible that the reef inlet itself creates some wave focusing which may worsen the impact on the stretch of shoreline corresponding with the location of the reef opening. In other words, while the coral reef maybe protecting some parts of the coastline, it maybe simultaneously focusing tsunami energy towards specific inlets or entrance channels as observed in some of the simulations of the study.

While tsunami wave elevations are found to be significant in the two communities under study, considering the proximity of the Fiji Islands to two major subductions zones and the size of the events modeled here, the geographical location of Fiji, lying on the overriding plate of both the New Hebrides and Tonga-Kermadec subduction zones is most likely providing some degree of protection from the brunt of the waves, as tsunamis generated in subduction zones tend to have a significant amount of their energy reflected from the overriding plate in the direction of the subducting one.

This study provides one more step toward assessing the effect of tsunamigenic earthquakes on Fiji. However, this study does not consider tsunamis generated by volcanic eruptions or tsunamigenic landslides that may occur due to earthquakes or volcanic activity as was the case during the 1953 event off the coast of Suva.
It is also important to remember that in order for tsunami dynamics river to be modeled accurately up the Sigatoka, precise bathymetric data and reliable elevation values of river banks and Sigatoka river mouth are necessary.

The final inundation map produced in this report does not respond to a single tsunami event, but it is the combination of a series of wort-case scenario events considered for the study site. However, data from each individual worst-case scenario has also been presented and provided. These data are associated with potential scenarios and could at some point materialize and provide validation data for the study.

Model output products are provided (in GIS-readable TIFF format). They include the maximum wave amplitudes, the maximum flow depths as well as composite over all sources for maximum wave amplitude/tsunami height and flow depth.